\documentclass[12pt]{article}
\pdfoutput=1
\usepackage{graphicx} 
\usepackage{amsmath,amsfonts,amssymb,stmaryrd,latexsym}
\usepackage{url,subcaption,comment}
\usepackage{hyperref}
\usepackage{mathtools}
\usepackage{xcolor} 

\hypersetup{
    colorlinks=true,       
    linkcolor=purple,    
    citecolor=blue,        
    filecolor=magenta,      
    urlcolor=black           
}
\usepackage{cite}
\usepackage{dcolumn}
\newcommand{\be}{\begin{equation}}
\newcommand{\ee}{\end{equation}}
\newcommand{\bea}{\begin{eqnarray}}
\newcommand{\eea}{\end{eqnarray}}
\newcommand{\bear}{\begin{eqnarray}}
\newcommand{\eear}{\end{eqnarray}}
\newcommand{\ba}{\begin{array}}
\newcommand{\ea}{\end{array}}

\newcommand{\eg}{{\it e.g.}}
\newcommand{\ie}{{\it i.e.}}
\newcommand{\etc}{{\it etc.}}
\renewcommand{\arraystretch}{1.3}

\begin{document}

\baselineskip=18pt \pagestyle{plain} \setcounter{page}{1}

\vspace*{-1cm}

\noindent \makebox[11.9cm][l]{\small \hspace*{-.2cm} }{\small Fermilab-Pub-20-030-T}  \\  [-1mm]

\begin{center}

{\Large \bf  Chiral Abelian gauge theories with few  fermions}    \\ [9mm]

{\normalsize \bf Davi B.~Costa$^\diamond$, Bogdan A. Dobrescu$^\star$ and Patrick J. Fox$^\star$ \\ [3mm]
{\small \it $\diamond$ University of Sao Paulo, SP 05508-090, Brazil \\
$\star$Theoretical Physics Department, Fermilab, Batavia, IL 60510, USA     }}\\

\center{January 30, 2020}
\end{center}

\vspace*{0.2cm}

\begin{abstract}  \normalsize
We construct chiral theories with the smallest number $n_\chi$ of Weyl fermions that form an anomaly-free set under various Abelian gauge groups.
For the $U(1)$ group, where $n_\chi = 5$,  we show that the general solution to the anomaly equations
is a set of charges given by cubic polynomials in three integer parameters.
For the $U(1) \times U(1)$ gauge group we find $n_\chi = 6$, and derive the general solution to the anomaly equations, in terms of 6 parameters.
For $U(1) \times  U(1) \times U(1)$ 
we show that $n_\chi = 8$, and present some families of solutions. 
These chiral gauge theories have potential applications to dark matter models, right-handed neutrino interactions, and other extensions of the Standard Model. As an example, we present a simple dark sector with a natural mass hierarchy between three dark matter components.
\end{abstract}

\vspace*{0.7cm}

\newpage

\hypersetup{linktocpage} 
\tableofcontents
\hypersetup{linkcolor=purple}   

\makeatletter
\let\toc@pre\relax
\let\toc@post\relax
\makeatother



\bigskip   \bigskip    \bigskip   
\section{Introduction} \setcounter{equation}{0}
\label{sec:intro}

Gauge theories with chiral fermions have remarkable properties. The fermion masses can be generated only if the 
chiral gauge group is spontaneously broken.  This provides an upper limit on the fermion masses, given approximately 
by the vacuum expectation value that breaks the gauge symmetry.  
By contrast,  vectorlike fermions  (\ie, fermions whose masses are gauge invariant) 
are generically expected to be as heavy as the largest mass scale present in the theory.

The gauge charges of chiral fermions 
are strongly constrained by the requirement that the gauge symmetry
is anomaly-free, \ie, it is preserved at the quantum level  \cite{Bardeen:1969md}. 
The only potential loophole would be the cancellation of the anomalies by an axion which interacts with the 
gauge bosons via a dimension-5 coupling. 
Nevertheless, if the scale of gauge symmetry breaking is much smaller than the Planck scale, as is the case in experimentally-accessible 
models, then the dimension-5 coupling should arise from integrating out some chiral fermions of mass comparable to the 
gauge boson masses \cite{Preskill:1990fr}. Thus, the charges of all fermions should still form an anomaly-free set.
In other words, the fermion charges must be solutions to certain anomaly equations.
Generically, these are cubic equations because the gauge anomalies in four spacetime dimensions arise from triangle diagrams with the 
gauge bosons on the  external lines and  the fermions running in the loop.

In Abelian gauge theories there are no interactions involving only gauge bosons, so the gauge couplings grow with the energy scale.
Thus, these theories are well-behaved at high energies only if they can be embedded in some non-Abelian structures.
A necessary condition for that embedding is that the Abelian charges are commensurate, which implies that 
they can be taken as integers upon an appropriate redefinition of the gauge couplings. A discussion of some additional arguments 
leading to the same conclusion can be found in \cite{Banks:2010zn}.
Thus, the cubic anomaly equations must be solved with integer variables.  
This is a vexing mathematical challenge given that it is impossible to construct an universal algorithm for solving such Diophantine equations \cite{Hilbert}, and that general solutions are not known in most nontrivial cases \cite{Hardy,Dickson}.

For the simplest Abelian gauge group, namely a single $U(1)$, the most general solution to the anomaly equations 
has been found in \cite{Costa:2019zzy}. 
That result shows that the most general $U(1)$ charges of  $n$ Weyl fermions are given by certain quartic polynomials in 
$n-2$ integer parameters.

No such general solution is known for larger Abelian gauge symmetries, which are products of $U(1)$ groups.
The numerical search for solutions to the anomaly equations is useful only when 
the range of fermion charges is very small \cite{Allanach:2018vjg}. To see that, note that in the case of $m$ $U(1)$'s with $n$ fermions 
carrying integer charges ranging from $-z$ to $+z$, the search needs to cover a number of configurations of the order of $(2 z +1)^{m \, n}$.
For $z = O(10)$, $m= 2$ and $n \ge 6$, that number is already above $10^{15}$.

A method useful in the case of arbitrary gauge groups, including product groups, is the construction of an anomaly-free chiral set starting from 
any anomalous set of fermions and adding some new fermions of certain charges \cite{Batra:2005rh}.
Another method is to start from some non-Abelian gauge group with a known anomaly-free representation and analyze its $U(1)$ subgroups \cite{deGouvea:2015pea}.
It is not yet clear whether algebraic geometry methods \cite{Lu:2019rro} may lead to general solutions for multiple $U(1)$ groups.

In this paper we analyze the Abelian gauge theories with the smallest number $n_\chi$ of chiral fermions.
For a single $U(1)$, it is known that $n_\chi =5$  \cite{Babu:2003is,Davoudiasl:2005ks}, so that  the minimal chiral theory
includes five Weyl fermions with charges depending on  three
integer parameters \cite{Costa:2019zzy}. We show that the general solution to the anomaly equations simplifies in this case,
being given by cubic polynomials in the three parameters. This parametrization allows us to prove that the set of solutions is infinite,
and that the relative signs of all five charges are fixed.

For the  $U(1) \times U(1) $ gauge theories we find that the smallest number of chiral fermions is $n_\chi =6$,
and we construct the general solution to the anomaly equations for this chiral theory, in terms of six integer parameters.
For the  $U(1) \times U(1) \times U(1)$  gauge group, we show that chiral theories exist only for $n \ge 8$ fermions, 
and we identify some anomaly-free chiral sets of charges which depend on three integer parameters.

The chiral gauge theories constructed here have potential applications to various model building efforts related to 
gauge extensions of the Standard Model \cite{Appelquist:2002mw,Carena:2004xs},  flavor structures \cite{Bonnefoy:2019lsn}, 
 right-handed neutrino interactions \cite{Babu:2003is,Davoudiasl:2005ks,Appelquist:2002mw,Sayre:2005yh,Babu:2004mj,Heeck:2012bz}, dark matter sectors \cite{Nakayama:2011dj,Heeck:2012bz},  and other phenomena.  
Chiral models have great advantages over models with vectorlike fermions. The chiral fermions 
are ensured to be much lighter than the Planck scale, and furthermore large mass hierarchies between fermion flavors 
naturally arise in chiral models.
As an example, we show that even one of the simplest $U(1)\times U(1)$ chiral models obtained here 
is a viable dark sector with complex phenomenological implications, which includes three dark matter components  
with hierarchical masses.

In Section \ref{sec:general} we display the anomaly equations for arbitrary Abelian gauge theories, summarize the most general solution for a single $U(1)$,
and define some key concepts (primitive solutions, canonical order for chiral sets of charges, \etc) used in later Sections.
The  $U(1)$ gauge theory with 5 fermions is analyzed in Section  \ref{sec:U1}.
The general solution to the $U(1) \times U(1) $ gauge theory with 6 fermions is derived in Section  \ref{sec:2U1}.
The smallest number of chiral fermions charged under three $U(1)$ groups is derived in Section \ref{sec:3U1}, along with some families of solutions.
Our conclusions are collected in Section  \ref{sec:conclusions}.

\bigskip
\section{Anomaly equations for Abelian gauge groups}
\setcounter{equation}{0}\label{sec:general}
 
We study Abelian gauge theories with a number $m$ of $U(1)$ groups whose charges are carried by  $n$ Weyl fermions.
The charges of the $i$th fermion, for any $1\leq i \leq n$, under the $U(1)^m \equiv U(1)_1 \times U(1)_2  \times  ...   \times  U(1)_m$  gauge group are 
 labelled by $[z_{1i},  z_{2i}, ... , z_{mi} ]$,  where all the Weyl fermions are taken to have the same chirality (\eg, left-handed).
 The set of charges of the $n$ fermions under the $U(1)_j$ gauge group is labelled by  
 \be
 \{ \vec z_j \} \equiv \{ z_{j1},  z_{j2}, ... , z_{jn} \}   \;\; {\rm for} \;\;  1\leq j \leq m    ~~.  
 \label{eq:set1}
\ee
As the charges within the $ \{ \vec z_j \} $ set are commensurate, we can choose all $z_{ji}$ charges to be integers  
by rescaling the $g_j$ gauge coupling of $U(1)_j$, for $1\leq j \leq m $.

The fermion charges must satisfy several anomaly equations.  
For each $U(1)$ group there is an $[U(1)]^3$ anomaly equation, due to triangle diagrams with three external gauge bosons \cite{Bardeen:1969md},
and a gravitational-$U(1)$ anomaly equation, due to diagrams with one external gauge boson and two external gravitons \cite{Eguchi:1980jx}. 
The first type of equation is cubic in charges, while the second one is linear in charges:
\bear 
\sum_{i=1}^n  z_{ji}^3 = 0 ~~,     \;\; {\rm for} \;\;  1\leq j \leq m   ~~,
\label{eq:cubic}
\\ [3mm]
 \sum_{i=1}^n  z_{ji} = 0 ~~,      \;\; {\rm for} \;\;  1\leq j \leq m  ~~.
 \label{eq:linear}
\eear 
Eliminating $z_{jn}$ from  Eq.~(\ref{eq:linear}), we can rewrite Eq.~(\ref{eq:cubic}) in the form ``sum of the cubes equals the cube of the sum",
with $n-1$ integer variables:
\be
\sum_{i=1}^{n-1}  z_{ji}^3 =  \left( \sum_{i=1}^{n-1}  z_{ji}  \right)^{\! 3} ~~,     \;\; {\rm for} \;\;  1\leq j \leq m   ~~.
\label{eq:cubes}
\ee

There are also anomaly equations involving each pair of $U(1)$ groups, due to triangle diagrams with two external gauge bosons
of one kind and a single external gauge boson of the other type. These $[U(1)_j]^2 U(1)_{j'}$ and $U(1)_j [U(1)_{j'}]^2$ 
anomaly equations (with $j \neq j'$) are given by
\bear 
\sum_{i=1}^n  z_{ji}^2 \, z_{j' i} = 0 ~~,     \;\; {\rm for} \;\;  1\leq j,j' \leq m   ~~,
\label{eq:21}
\\ [3mm]
 \sum_{i=1}^n  z_{ji} \, z_{j' i}^2 = 0 ~~,      \;\; {\rm for} \;\;  1\leq j,j' \leq m   ~~.
 \label{eq:12}
\eear 
Finally, there are $U(1)_j U(1)_{j'} U(1)_{j''}$  anomaly equations (these are new only when no two indices are equal, which requires $m \ge 3$), 
due to diagrams with three different external gauge bosons:
\be
\sum_{i=1}^n  z_{ji} \, z_{j' i} \, z_{j'' i} = 0 ~~,     \;\; {\rm for} \;\;  1\leq j,j' , j'' \leq m   ~~.
\label{eq:111}
\ee
Altogether, the number of anomaly equations 
is
\begin{align}
{\cal N}_{\rm eq} = 2m+2\binom{m}{2}+\binom{m}{3}=\frac{m}{6}  \left( m^2+3m+8 \right)    ~~,
\label{eq:neq}
\end{align}
where $2m$ is due to the $U(1)$ and $[U(1)]^3$ anomalies, $2\binom{m}{2}$ is the number of $U(1)_j  [U(1)_{j'} ]^2$ mixed anomalies, 
and $\binom{m}{3}$ is due to the $U(1)_j U(1)_{j'} U(1)_{j''}$  anomalies.

For each $U(1)_j$ gauge group, the most general solution to the $[U(1)_j]^3$ anomaly equation (\ref{eq:cubic}) and the  gravitational-$U(1)_j$ anomaly equation (\ref{eq:linear}), depends on $n-2$ integer parameters.  
That general $U(1)_j$ solution was found in \cite{Costa:2019zzy},  and can be written as the ``merger" of two vectorlike sets of charges. The merger operation\footnote{A geometric interpretation of the merger is given in \cite{Allanach:2019gwp}.}, labelled by  $\oplus $, acts on two sets of charges of equal length, similar to those of Eq.~(\ref{eq:set1}), and  is defined by  
\be
\hspace*{-0.2cm}
\{ \vec{x}  \,  \}   \oplus   \{\vec{y}  \,  \}   \equiv   \left(  \sum_{i=1}^n x_i y_i^2 \right)  \{ \vec{x}  \,  \}  -  \left(  \sum_{i=1}^n x_i^2  y_i \right)   \{\vec{y}  \,  \}  ~~.
\label{eq:xyzset}
\ee
For even $n$, the general $U(1)$ solution (here we drop the $j$ index to simplify the notation) 
is the merger of two vectorlike sets of charges, and takes the form
\be
 \{   \ell_1, k_1, ... \, , k_{n/2-1} , -  \ell_1 , -k_1, ... \, , - k_{n/2-1} \}  
  \oplus 
\{  0, 0,  \ell_1, ... \, ,  \ell_{n/2-1} , -  \ell_1, ... \, , -  \ell_{n/2-1}  \}   ~~,
\label{eq:even-merger}
\ee
where $\ell_1, ... , \ell_{n/2-1}$ and $k_1, ... , k_{n/2-1} $ are integer parameters.
For odd $n$, the general $U(1)$ solution can be written as
\be
 \{0, k_1, ... \,  , k_{(n-1)/2 }, -k_1, ... \,  , -k_{(n-1)/2 } \} 
   \oplus   
 \{ \ell_1, ...  \, , \ell_{(n-3)/2 }, k_1, 0, -\ell_1, ... \, , - \ell_{(n-3)/2 }, -k_1 \}  
 \label{eq:merger-odd}
\ee
with the $n-2$ integer parameters labelled in this case by $k_1, ... \,  , k_{(n-1)/2 }$ and $\ell_1, ... \, , \ell_{(n-3)/2 }$.
Notice that the above forms of the general $U(1)$ solution imply that charges are given by polynomials in the integer parameters,
which collectively are quartic. 
We use the phrase ``general solution" not as a reminder that it applies to any number of fermions, but rather  to emphasize that {\it any}
chiral set of charges that satisfy the two $U(1)$ anomaly equations for fixed $n$ corresponds to a particular choice of the integer parameters. 

It is worth mentioning that there are some classes of trivial solutions to the $U(1)$ anomaly equations  that are immediately obtained from the general solution
(\ref{eq:even-merger}) or (\ref{eq:merger-odd}) by changing the normalization of all charges, 
or by making permutations among the $n$ charges. Furthermore, vectorlike solutions are trivial, and are completely 
covered, up to permutations among the $n$ charges, by the first set of either (\ref{eq:even-merger}) for even $n$, or 
(\ref{eq:merger-odd}) for odd $n$.

Even though the general solution
(\ref{eq:even-merger}) or (\ref{eq:merger-odd}) 
applies 
only to individual $U(1)$ groups, it is also useful for the case of
product groups (see Sections \ref{sec:2U1} and \ref{sec:3U1}). For example, 
to find anomaly-free sets of fermions charged under  $U(1)_1  \times  ...   \times  U(1)_m$,
one could first write the general $U(1)_j$ solution for each $j$, and then try to solve 
the mixed anomaly equations (\ref{eq:21})-(\ref{eq:111}) to reduce the number of integer parameters.

If two of the Weyl fermions form a vectorlike pair,
\ie, their $U(1)_1 \times  ...   \times  U(1)_m$
charges satisfy $[ z_{1i},  z_{2i}, ... , z_{mi} ] = - [z_{1i'},  z_{2i'}, ... , z_{mi'} ] $,
then that pair can acquire a large mass and  its contributions to the anomaly equations vanish, 
so the problem reduces to $n-2$ Weyl fermions.
Thus, it is sufficient to consider chiral sets of fermions, which means that the following condition 
must be satisfied:
\be
[z_{1i},  z_{2i}, ... , z_{mi} ] \neq - [z_{1i'},  z_{2i'}, ... , z_{mi'} ]
 \;\;\; {\rm for}  \;\,  {\rm all}  \;\;\; 1 \leq i,i' \leq n  ~~~,
\label{eq:chiral}
\ee
where the above inequality does not preclude $z_{ji} = - z_{ji'}$ for some or most  (but not all) values of $j$ in the $1 \leq j \leq m$ interval.
That is, two Weyl fermions may form a vectorlike pair with respect to some of the $U(1)_j$ subgroups and still be chiral with respect to the whole 
$U(1)^m$ gauge group.

As all anomaly equations (\ref{eq:cubic})-(\ref{eq:111}) are homogeneous in the charges of each $U(1)$ group, 
for any chiral set of integer charges which satisfies them there is an infinite set of solutions where all the charges 
under any $U(1)$ group are multiplied by an arbitrary integer. It is thus sufficient to consider sets  of charges 
which are  coprime
with respect to each $U(1)$, \ie, the greatest common divisor of the $n$ charges under each $U(1)$ is 1. 
We will refer to such a set as coprime.

Thus a solution to the anomaly equations is nontrivial if it is chiral and coprime. 
For large enough $n$ the solutions include composite sets, which are formed of two or more subsets that independently satisfy 
the anomaly equations. 
Extending the terminology of  \cite{Costa:2019zzy} to multiple $U(1)$'s, we define ``primitive" solutions  as anomaly-free chiral sets which are coprime but not composite.

Each  primitive solution belongs to a family of $n! \, 2^m $  primitive solutions which differ only by a reordering of the fermions or by sign flips of all charges 
under any of the $U(1)$ groups.
To avoid this trivial multiplicity, we define the ``canonical form" for a set of $U(1)_1 \times ... \times U(1)_m$ charges as the 
ordering according to the decreasing absolute values of the charges under $U(1)_1$, and with the first 
fermion (the one with the $U(1)_1$ charge of largest absolute value)  
having only positive charges.
In the case of a single $U(1)$, this is identical to the  canonical form defined in  \cite{Costa:2019zzy}.
To be more explicit, let us label the set of $U(1)_1\times ... \times U(1)_m$ charges by  
\be
\left[ \vec z_1 , \vec z_2 , ... , \vec z_m \right] \equiv
\left[ \!
\begin{array}{c|c|c|c}    
z_{11} & z_{21} & ... & z_{m 1} \\
z_{12} & z_{22}  & ... & z_{m 2} \\
\vdots & \vdots   &  & \vdots   \\
z_{1n} & z_{2n}  & ... & z_{m n} \\
\end{array}  
\! \right]  ~~.
\ee
This set is in canonical form if and only if 
\bear
&& z_{11} \ge |z_{1 2}| \ge  ...  \ge |z_{1 n}|     ~~~,    
\nonumber \\ [2mm]
&& z_{i 1} \ge 0  \;\;\;  {\rm  for \,\,  all}  \;\; 1 \leq i \leq m ~~.
\label{eq:canonical}
\eear

Note that even for $m \ge 2$ primitive solutions in the canonical form belong to families of $m!$ sets related by interchanges of the 
$U(1)$ groups. That additional trivial multiplicity can also be eliminated by imposing the prescription that  $z_{11}$ has the largest 
absolute value of all $z_{ji}$ charges, with additional prescriptions in the cases where there are two or more such charges.
However, for the purpose of this article it is not necessary to impose these prescriptions.
 
\bigskip
\section{$U(1)$ gauge group with 5 Weyl fermions}
 \setcounter{equation}{0}   \label{sec:U1}

There are no chiral anomaly-free sets with four or fewer fermions charged under a $U(1)$ gauge symmetry \cite{Babu:2003is,Davoudiasl:2005ks}.
The smallest number of Weyl fermions that allows chiral solutions to the anomaly equations (\ref{eq:cubic}) and (\ref{eq:linear}) 
is $n_\chi=5$. Several chiral solutions have been found numerically \cite{Babu:2003is, Davoudiasl:2005ks, Sayre:2005yh, Batra:2005rh}, with the lowest lying ones being
$\{ 9, -8, -7, 5, 1\}$ and  $\{ 10, -9, -7, 4,2 \}$.

In the case of a single $U(1)$ gauge symmetry, we simplify the notation for fermion charges introduced in Section \ref{sec:general} 
by dropping the index that labels the gauge group: $z_{1i} \equiv z_i$, for $1\leq j \leq n$.
The anomaly equation of the type ``sum of the cubes equals the cube of the sum",  Eq.~(\ref{eq:cubes})
for $n=5$  (and $m=1$, \ie, a single $U(1)$ group)  is 
\be
z_1^3 + z_2^3 + z_3^3 + z_4^3  =  \left( z_1 + z_2 + z_3 + z_4  \right)^3 ~~,
\label{eq:cubes5}
\ee
while the fifth charge is given by
\be
 z_5 = - \left( z_1 +z_2 +z_3 +z_4  \right)  ~~.
\label{eq:linear5}
\ee
The chiral conditions  (\ref{eq:chiral})  become
\be
z_i +  z_{i'}\neq 0  \;\;\; {\rm  for \;\;  any \;\; }  1 \leq i, i' \leq 5  ~~.
\label{eq:chiralU1}
\ee
Adapting  Eq.~(\ref{eq:canonical}) to the case of $m=1$, $n=5$, a chiral set  $ \{z_1, ... , z_5 \} $ 
is in the canonical form (as defined in  \cite{Costa:2019zzy}) if it satisfies
\be
z_1 \ge |z_2| \ge  |z_3| \ge |z_4| \ge |z_5| \ge 1    ~~~.     
\label{eq:canonical5general}
\ee

Next we derive some generic properties of the anomaly-free chiral sets of 5 Weyl fermions.
Later in this Section we will present and analyze the general solution to the anomaly equations for a single $U(1)$ with $n=5$.

\subsection{Properties of the 5-charge chiral sets}

We first point out that all chiral solutions to Eqs.~(\ref{eq:cubes5})  and (\ref{eq:linear5}), when arranged in canonical form,  have a unique signature.  
To prove this, consider the following equivalent form of Eq.~(\ref{eq:cubes5}) for $n=5$, obtained using Eq.~(\ref{eq:linear5}):      
 \be
z_1 z_5 \left( z_1 +  z_5   \right) = - \left( z_2 +  z_3   \right) \left( z_2 +  z_4   \right) \left( z_3 +  z_4   \right)   ~~.
\label{eq:cubic3}
\ee

Using the canonical ordering (\ref{eq:canonical}) and imposing the chiral conditions  (\ref{eq:chiralU1}) 
we find $z_1  \left( z_1 +  z_5   \right) > 0$ and $\left( z_2 +  z_3   \right) \left( z_2 +  z_4   \right) > 0$.  
Given that  $z_3 +  z_4$ has the same sign as $z_3$,  Eq.~(\ref{eq:cubic3}) implies
that $z_3$ and $z_5$ have opposite sign.
As the anomaly equations  (\ref{eq:cubes5})  and (\ref{eq:linear5}) are symmetric under interchange of any pair of charges, we 
can rewrite  Eq.~(\ref{eq:cubic3}) by interchanging $z_2$ and $z_5$. Following the same steps as above, it follows that 
$z_2$ and $z_4$ have opposite sign.
Analogously, interchanging $z_3$ and $z_5$ in Eq.~(\ref{eq:cubic3})  implies that $z_3$ and $z_4$ have opposite sign. 
In addition, $z_1$ and $z_2$ must have opposite sign  in order for the sum of five charges to vanish. 
Thus, the only pattern of signs for charges that solves the anomaly  equations is 
\be
\{ +, -, -, +, +\}     
\label{eq:signature}
\ee
when the charges are ordered in canonical form. 
The uniqueness  of the above signature is a special property of the $n = 5$ case. For $n=6$ there are two possible signatures \cite{Costa:2019zzy}, and for larger $n$ the number of signatures grows quickly.

We next prove that no two charges are  equal  in the chiral set for $n = 5$.
Due to the signature established above, it is sufficient to show that $z_2 \neq z_3$ and $z_4 \neq z_5$. 
If $z_2 = z_3$ were true, then Eqs.~(\ref{eq:cubic3}) and (\ref{eq:linear5}) would lead to the following quadratic equation in $z_4$:
\be
(z_1 + 2 z_2) \, z_4^2 +(z_1+ 2 z_2)^2 \, z_4 +  2 z_2 (z_1 + z_2)^2 = 0  ~~.
\label{eq:z4}
\ee
The discriminant $\Delta$ of this equation can be written as the difference of two perfect squares:
\be
\Delta = \left(z_1^2 - 2 z_2^2\right)^2 - \left(2 z_2 (z_1 + z_2) \rule{0mm} {3.8mm}\right)^2  ~~.
\ee
A necessary condition for Eq.~(\ref{eq:z4}) to have an integer solution is that $ \Delta$  is a perfect square.
Thus, the three integers $2 z_2 (z_1 + z_2) $, $\Delta$ and $z_1^2 - 2 z_2^2$ must form a Pythagorean triple, which implies (see, \eg, pages 245--247 of \cite{Hardy}) 
that there must exist two integers $m$ and $m'$ such that 
$z_2 (z_1 + z_2) =   m \, m' $ and $z_1^2 - 2 z_2^2 =m^2 + m^{\prime \, 2} $.
These equations give $z_1 ( z_1 + 2 z_2) = (m + m')^2 > 0$, which is in contradiction with the inequalities 
$z_1 > 0$ and $ z_1 + 2 z_2 =-z_4 -z_5 <0$. 
Thus, we have shown that $z_2 < z_3$. 

Our proof that $z_4 \neq z_5$ is more elaborate. 
If $z_4 = z_5$ were true,  first we follow the same steps as above with 
$z_2 \leftrightarrow z_4$ and $z_1 \leftrightarrow z_3$, which leads to $z_3 ( z_3 + 2 z_4) > 0$, implying $-z_3 > 2 z_4$.
Second, Eqs.~(\ref{eq:cubic3}) and (\ref{eq:linear5}) with  $z_4 = z_5$ 
imply that the following quadratic equation in $r = -z_1/z_3 > 1$ would need to be satisfied:
\be
r^2 = (1 - a) \, r +  \frac{a ( 2-a)^2}{4(1-a)}  ~~,
\label{eq:z1r}
\ee
where the parameter introduced here is $0 <  a = - 2z_4/z_3 < 1 $. It is straightforward to show that Eq.~(\ref{eq:z1r}) has no solutions with  $r > 1$,
and consequently the assumption $z_4 =  z_5$ is false.
This completes the proof that the chiral set does not include two charges which are  equal.
 Combining this result with the signature (\ref{eq:signature}), 
we find that the charges of an anomaly-free chiral set in the canonical form satisfy a remarkable constraint:
\be
z_1 >  -z_2 >  - z_3 > z_4 > z_5 \ge 1  ~~. 
\label{eq:canonical5}
\ee

We now turn to the question of whether the number of primitive    
solutions to Eqs.~(\ref{eq:cubes5}) and (\ref{eq:linear5})  is infinite. To that end we construct a 
1-parameter family of solutions:
\be
 \left\{
 (2k + 1 )^2 ,\; -  k  (4 k + 3 )   ,\; - 2(k + 1 )^2  ,\;  2 k^2  - 1    ,\;  3 k + 2   \rule{0mm}{4mm}
\right\}    ~,
\label{eq:infinite1}
\ee
where $k \in \mathbb{Z}$ and $k \ge 1$.
Notice that this chiral set is already in the canonical form for $k \ge  3$.
We still need to find for which values of $k$ this set is coprime.
Let us concentrate on the  first and third charges of this set, namely  $( 2k + 1)^2 $ and  $-2(k + 1 )^2$.
If these have a common divisor $z_\star \ge 1 $, then $2k + 1 \equiv 0$ (mod $z_\star$) and $k + 1 \equiv 0$ (mod $z_\star$), which implies 
$1 \equiv 0$ (mod $z_\star$) so that $z_\star=1$.    
It follows that   the  first and third charges are always coprime, so that the whole set is coprime for 
any $k \ge 1$.
Thus,  the chiral set (\ref{eq:infinite1}) represents a countably infinite family of primitive  solutions.

It is worth noting that the set (\ref{eq:infinite1}) for $k\gg 1$ has two charges of order $\pm 4 k^2$, two charges of order $\pm 2k^2$, and one of order $3k$.
Therefore, in the large $k$ limit this chiral set is a small (but essential) perturbation away from a vectorlike set.

\subsection{General solution to the 5-charge anomaly equations}  

The most general nontrivial solution to the $U(1)$ anomaly equations (\ref{eq:cubic}) and (\ref{eq:linear}) for an odd number of 
Weyl fermions is given by the merger shown in Eq.~(\ref{eq:merger-odd}), as proved in  \cite{Costa:2019zzy}. 
Only the chiral solutions are nontrivial, as all vectorlike ones are
obtained from either set used in (\ref{eq:merger-odd}) and depend on $(n-1)/2$ parameters.

In the case of the smallest number of fermions, $n_\chi=5$,
the anomaly equations take the form (\ref{eq:cubes5}) and (\ref{eq:linear5}), and the general solution becomes
\be
 \{ \vec z \, \}  = \{0, k_1, k_2, -k_1, -k_2  \} 
   \oplus   
 \{ \ell, k_1, 0, -\ell,  -k_1 \}    ~~.
 \label{eq:merger-5}
\ee
Given that there are 5 charges and two equations, the general solution depends on 3 integer parameters labelled 
in this case by $k_1, k_2$ and $\ell$.

As explained in Section \ref{sec:general}, the general $U(1)$ solution gives the $n$ charges as quartic polynomials in the 
integer parameters. We point out that in the $n=5$ case a simplification occurs: the 5 charges arising from  the merger (\ref{eq:merger-5})
are proportional to $k_1 \neq 0$,
so dividing  all charges by $k_1$ we find that the solution is {\it cubic} in the integer parameters. 
Explicitly, our 3-parameter solution to the $U(1)$ anomaly equations for $n=5$ charges is 
\bear
&& z_1 = \ell \left(k_2^2 - k_1^2 + k_1 \, \ell\right)   
\nonumber \\ [2mm]
&& z_2 = -k_1 (\ell - k_2) (k_2+\ell-k_1)
\nonumber \\ [2mm]
&& z_3 = -k_2 \left( \ell^2 - k_1^2 + k_1 \, k_2 \right)
\label{eq:5set}
\\ [2mm]
&& z_4 = (k_2-k_1)
   \left( k_1^2-\ell (k_1+k_2) \right)
\nonumber \\ [2mm]
&& z_5 =(k_1-\ell) \left(k_1^2- k_2 (k_1+ \ell) \right)~~.
\nonumber 
\eear
Notice that each charge is a homogeneous cubic polynomial in the three integer parameters, but as functions of individual parameters 
the charges are quadratic or in some cases linear.
It is straightforward, albeit tedious,  to check that this is an anomaly-free chiral set.

The necessary and sufficient condition for any 5-charge set to be chiral, \ie, to satisfy Eq.~(\ref{eq:chiral}), is that none of the charges is 0.
Thus, the above set is chiral  if and only if all of the following conditions are satisfied:
\bear
&&  \hspace*{-1cm}
 k_1,k_2,\ell \neq 0  \; {\rm , }  \;\;\;  k_1 \neq k_2\;  , \;\; k_2 \neq \ell \; , \;\;  \ell \neq k_2    \; {\rm , }  \;\;\;  k_2+\ell \neq k_1 ~~~,
\nonumber \\ [-2mm]  
\label{eq:chiral5}
 \\ [-2mm]  
&&    \hspace*{-2cm}
k_2^2 \neq k_1^2 - k_1 \, \ell     \;{\rm , }  \;\;\;   \ell^2 \neq k_1^2 - k_1 \, k_2   \; {\rm , }  \;\;\;   k_1^2  \neq \ell (k_1+k_2)    \; {\rm , }  \;\;\;     k_1^2  \neq k_2 (k_1+ \ell)~~.
\nonumber 
\eear

Any choice of the three integers $k_1,k_2,\ell$ corresponds to a 5-charge solution, which can also be obtained up to a reordering of charges, or an overall sign flip,
by a few other choices of $k_1,k_2,\ell$.  The simplest transformations of this type are
\be
 ( k_1,k_2,\ell )  \rightarrow  - (k_1,k_2,\ell )   \;\;\; {\rm and }  \;\;\;   z_i  \rightarrow  -z_i \,  \,  {\rm  for}  \;\;\;    i = 1, ... ,5  ~~,
\label{eq:signflip}
\ee
as well as 
\be
k_2 \leftrightarrow  \ell     \; , \;  z_1  \leftrightarrow  -z_3     \; , \;  z_2  \leftrightarrow  -z_2   \; , \;  z_4  \leftrightarrow  -z_5   ~~.
\label{eq:kl}
\ee
Consequently, no primitive solutions are removed when 
$k_1,k_2,\ell$ are restricted to 
\be
k_1 \ge 1 \;\;\; , \;\;\;  k_2 \le \ell - 1  ~~.
\label{eq:restrict}
\ee
A more complicated transformation that keeps the set invariant is 
\be
( k_1,k_2,\ell )  \to (k_2+  \ell - k_1   \; , \;  k_2-k_1    \; , \;  \ell-k_1  )  \; \; , \; \;  \{z_1, z_2, z_3, z_4, z_5 \}  \rightarrow   -\{ z_4, z_2, z_5, z_3, z_1 \}     ~~.
\label{eq:kl2}
\ee
This implies, in particular, that any solution corresponding to parameters of the type $( k_1 \, , \,  k_2 +k_1 \, , \, -k_2 +k_1 ) $ 
is equivalent to the solution arising from $( k_1 \, , \, k_2 \, , \,  -k_2 ) $.
There is also a more particular case of invariance, which seems related but does not arise from the transformation (\ref{eq:kl2}):
\be
( k_1, -k_1 ,\ell )  \to (   k_1, -\ell , -k_1 )    \; \; , \; \;  \{z_1, z_2, z_3, z_4, z_5 \}  \rightarrow   -\{ z_3, z_4, z_1, z_5, z_2 \}      ~~.
\label{eq:jj}
\ee

For certain values of $k_1,k_2,\ell$ the set (\ref{eq:5set}) represents a primitive solution (not necessarily in the canonical form), while for other values it is a primitive 
solution rescaled by a common integer factor. To reduce the latter category, it is useful to impose that the  three integers $( k_1,k_2,\ell )$ 
are coprime (not necessarily pairwise coprime).  A particular transformation 
that changes both the ordering of the charges within the set, and the greatest common divisor (GCD) of the set, while leaving invariant the 
primitive solution is the following:
\be
( k_1, k_2 ,2k_1 )  \to (  k_2, -k_1, 2 k_2 )    \; \; , \; \;  z_3  \leftrightarrow  z_5     \; , \;  z_2  \leftrightarrow  z_4        \; \; , \; \;    {\rm GCD }= k_1    \to      {\rm GCD} = k_2  ~~.
\label{eq:jjGCD}
\ee

In Table 1 we display the primitive solutions in canonical form with 
$z_1 \leq 32$. For each of the primitive solution we indicate a choice of $(k_1,k_2,\ell)$ that generates that solution. 
In many cases, retrieving the canonical form requires a reordering of charges within the set  (\ref{eq:5set}), the division by their GCD, and an overall sign flip.
Note that for each solution there exist other choices of the integers $(k_1,k_2,\ell)$,  as indicated by the invariances
shown in Eqs.~(\ref{eq:signflip})-(\ref{eq:jjGCD}).

\begin{table}[t!]
\begin{center}
\renewcommand{\arraystretch}{1.5}
\begin{tabular}{|c|c|c|}\hline  
 Primitive solution    &   $(k_1,k_2,\ell)$   &   GCD$(\vec{z} \, )$  
\\ \hline \hline
    $ \{ 9, -8, -7, 5, 1 \} $   &  $( 2, -1, 1 )$    &  1
\\ [1mm]    \hline
      $  \{ 10, -9, -7, 4, 2  \}    $    &  $(  1, -2, 2  )$    &  1
\\ [1mm]    \hline
      $  \{  20, -18, -17, 14, 1  \}   $     &  $( 3, -2, 1   )$    &  2
\\ [1mm]    \hline
     $   \{  22, -21, -12, 6, 5  \}   $     &  $( 2, -4, -1   )$    &  2
\\ [1mm]    \hline
    $    \{  25, -22, -18, 8, 7 \}   $     &  $(  2, -4, 1  )$    &  2
     \\ [1mm]    \hline
 \hspace*{5mm}        $  \{ 26, -22, -20, 9, 7  \}   $     \hspace*{5mm}     &   \hspace*{5mm}   $(  2, -3, 4  )$     \hspace*{5mm}   &  2
\\ [1mm]    \hline
     $   \{  26, -25, -14, 9, 4 \}    $    &  $( 1, -4, -2   )$    &   1
\\ [1mm]    \hline
      $  \{ 27, -25, -17, 8, 7  \}    $    &  $(  1, -4, 3  )$    &  2
\\ [1mm]    \hline
      $  \{ 27, -26, -14, 8, 5  \}    $    &  $( 2, 3, 4   )$    &  2
\\ [1mm]    \hline
      $  \{  28, -25, -23, 18, 2 \}    $    &  $( 3, -2, 4   )$    &  1
\\ [1mm]    \hline
      $  \{ 28, -26, -18, 11, 5  \}    $    &  $( 3, -1, 1   )$    &  1
\\ [1mm]    \hline
   \hspace*{5mm}     $  \{ 32, -27, -25, 13, 7  \}      \hspace*{5mm}   $    &   \hspace*{5mm} $(  2, -3, 1  )$   \hspace*{5mm}  &  1
\\ [1mm]     \hline
\end{tabular}
\caption{Solutions to the anomaly equations for 5 chiral fermions. Only 
primitive solutions in canonical form are shown, for $z_1 \leq  32$.  For each set of 5 charges we show a particular choice of the $(k_1,k_2,\ell)$ integer parameters 
and the greatest common divisor that generate the primitive solution from the general set (\ref{eq:5set}).   }
\label{table:1}
\end{center}
\end{table}

The proof that {\it all} chiral solutions for arbitrary $n$ can be generated by the merger operations
 of Eqs.~(\ref{eq:even-merger}) and (\ref{eq:merger-odd}) was given in \cite{Costa:2019zzy}  for an arbitrary number of fermions.
 It is instructive to adapt that proof of generality to the case of $n=5$. Consider an arbitrary chiral set of 5 charges, $\{ \vec{q}  \, \}$,
 which is a solution to the anomaly equations  (\ref{eq:cubes5}) and (\ref{eq:linear5}). We need to show that 
there exists three integers $(k_1,k_2,\ell)$ such that the set $\{ \vec{z}  \, \}$ of 5 charges given in (\ref{eq:5set}) 
is the same as $\{ \vec{q}  \, \}$ up to an overall rescaling by an integer.
To that end, we identify $(k_1,k_2,\ell)$ in terms of the $\{ \vec{q} \, \}$ components as follows:
\bear
&& k_1 = q_4 (q_1 + q_4)
\nonumber \\ [1mm] 
&& k_2 = - q_3  q_4
\label{eq:jklproof}
 \\ [1mm] 
&& \ell =  (q_2 + q_3) (q_2 + q_5)~~.
\nonumber 
\eear
Replacing these expressions in the set (\ref{eq:5set})  we obtain each of the five $z_i$ quantities as 
6th-degree polynomials in the $q_1, q_2, q_3$ and $q_4$ charges of the $\{ \vec{q} \, \}$ set (after using the linear equation to eliminate $q_5$). 
The next step is to eliminate the higher powers of one of the charges, for instance $z_4$, 
using the cubic equation (\ref{eq:cubes5}) written in terms of $q_i$'s,
\be
q_4^2 = -(q_1 + q_2 + q_3) q_4 - \frac{(q_1 + q_2) (q_1 + q_3) (q_2 + q_3) }{q_1 + q_2 + q_3} ~~.
\label{eq:zero}
\ee
The result of this lengthy computation is that the $\{ \vec{z} \, \}$ set is proportional to $\{ \vec{q} \, \}$:
\be
 \{ \vec{z} \, \} =c_\star \, \{ \vec{q} \, \}  ~~.
\ee
The final step is to notice that Eq.~(\ref{eq:zero}) implies 
\be
c_\star =  q_4 (q_2 +  q_5 )    \left (   (q_2 + q_3)^2   (q_2+  q_5) + q_4^2     (q_1+ q_4)  \rule{0mm}{4mm}\right) ~~,
\ee
which shows that $c_\star$ is an integer. 
Thus,  the set $\{ \vec{z} \, \}$ of Eq.~(\ref{eq:5set}) is the general solution to the anomaly equations for 5 chiral fermions.

The relation between $(k_1,k_2,\ell)$  and the  $\{ \vec{q} \, \}$ set shown in 
Eq.~(\ref{eq:jklproof}) indicates that there is a more restrictive range for the integer parameters which still spans the general 
solution. When $\{ \vec{q} \, \}$ is in the canonical form,   analogous to (\ref{eq:canonical5}), the parameters satisfy
\be
\ell > k_1 > k_2 \ge 1  ~~.
\label{eq:octant}
\ee
Consequently, this restriction is sufficient for the set  (\ref{eq:5set}), with appropriate reorderings and an overall integer rescaling, to 
generate all the anomaly-free chiral solutions. This restriction is also sufficient for the chirality conditions (\ref{eq:chiral5}) to be satisfied, and it 
implies $z_1 > -z_2 > z_4 \geq 1$  and $z_1 > -z_3 > z_4$.
The $z_5 > 0$ condition is satisfied if and only if 
\be
\ell > \frac{k_1^2}{k_2} - k_1   ~~.
\label{eq:z5p}
\ee
Thus, imposing (\ref{eq:octant}) and (\ref{eq:z5p}) brings  the set  (\ref{eq:5set}) almost to the canonical form:
 the only remaining conditions are  $z_4 > z_5$, which sets an upper limit on $\ell/k_1$ as a function of $k_1/k_2$, 
and $|z_2| > |z_3|$ which sets a lower limit on $\ell/k_1$ for a range of  $k_1/k_2$.
The restriction  (\ref{eq:octant}), however,  has  a drawback:  to generate various solutions with small charges
one needs large values of $(k_1,k_2,\ell)$, and thus the set $\{\vec{z} \, \}$  needs to be divided by a large GCD$(\vec{z} \, )$. 
By contrast, the range (\ref{eq:restrict}) works with lower absolute values of $(k_1,k_2,\ell)$ (see Table 1) at the cost of having multiple 
coprime choices for $(k_1,k_2,\ell)$ associated with a single solution.

\bigskip
\section{General solution for $U(1)\times U(1)$  with 6 fermions}
 \setcounter{equation}{0}   \label{sec:2U1}

We now turn to the smallest set of fermions charged under a $U(1)_1\times U(1)_2$ gauge group that is chiral and anomaly free.  
Using the notation introduced in Section~\ref{sec:general}, 
the charges of $n$ Weyl fermions under the two $U(1)$  groups are $\{ \vec z_{1} \}$ and $\{ \vec z_{2} \}$, respectively. We impose that $\vec z_{1}$ is not parallel to $\vec z_{2}$ --- otherwise there is a field redefinition which removes couplings to one of the gauge bosons.  
If the whole set of $U(1)_1\times U(1)_2$ charges  
\be
\left[ \vec z_1 , \vec z_2 \right] \equiv
\left[ \!
\begin{array}{c|c} 
z_{11} & z_{21} \\
z_{12} & z_{22} \\
\vdots & \vdots  \\
z_{1n} & z_{2n}
\end{array} \!  \right] ~~,
\ee 
satisfies the anomaly equations (\ref{eq:cubic}), (\ref{eq:linear}), (\ref{eq:21}) and (\ref{eq:12}) for $m=2$, then any linear combination of the two sets of $U(1)$ charges,
\begin{equation}
 [ \vec z^{\; \prime}_1 , \vec z^{\; \prime}_2] = \left[ \kappa_1 \vec z_{1} + \lambda_1 \vec z_{2} \; , \,   \kappa_2 \vec z_{1} + \lambda_2 \vec z_{2} \,  \right]~,
 \label{eq:linear-comb-gen}
 \end{equation}
will also satisfy the anomaly equations.  
Conversely, if $\{ \vec z_{1} \}$ and $\{ \vec z_{2} \}$ are separately anomaly-free but do not satisfy the mixed $[U(1)_1]^2 U(1)_{2}$ and $U(1)_1 [U(1)_{2}]^2$ anomaly equations, then no linear combination will remove the mixed anomalies while preserving the $[U(1)]^3$ and gravitational-$U(1)$ anomaly equations.  To keep the charges integer valued we take  $\kappa_{1,2}, \lambda_{1,2} \in \mathbb{Z}$.  While the two theories described by fermions with charges $[\vec z_1, \vec z_2 ]$ and $[  \vec z^{\; \prime}_1 , \vec z^{\; \prime}_2 ]$ are related by the above linear transformations, when the two vector bosons have different masses, they are not equivalent and may actually have very different properties.

As discussed in Section 3, under a single $U(1)$ the smallest anomaly-free chiral theory requires 5 fermions. Is this still true for two $U(1)$'s?  Using the two linear transformations  (\ref{eq:linear-comb-gen}) we can transform to a set of charges $[ \vec{z}_1,\vec{z}_2 ]$ such that each of $\{\vec{z}_1\}$ and $\{\vec{z}_2\}$ have (at least) one of their charges equal to zero, which without loss of generality we take to be $z_{11} = z_{22} =0$. But the gravitational-$U(1)$ and $[U(1)]^3$ anomaly equations imply that the only anomaly-free solutions for 4 fermions have two pairs of vectorlike fermions \cite{Babu:2003is,Davoudiasl:2005ks}. There are two possible arrangements of these charges across the two groups.  Using the $[U(1)_1]^2 U(1)_{2}$ and $U(1)_1 [U(1)_{2}]^2$ anomaly equations, it is straightforward to show that for both cases the only solutions are vectorlike. 
Note that the linear transformations (\ref{eq:linear-comb-gen}) cannot transform a vectorlike solution into a chiral one, or vice-versa.
Thus, the minimum number of fermions such that $U(1)\times U(1)$ is chiral and anomaly-free is $n_\chi\ge 6$.

As we will show, by explicit construction, this minimum number of fermions is $n_\chi=6$. In Sections \ref{sec:chi-chi}--\ref{sec:dm}
we analyze the case where the set of charges $\{ \vec z_{1} \}$  under the first $U(1)$ is chiral, leaving the 
case where $\{ \vec z_{1} \}$ is vectorlike for Section \ref{sec:doublyvec}.

\subsection{Doubly-chiral and chiral-vectorlike sets under $U(1)\times U(1)$}
\label{sec:chi-chi}

Any anomaly-free  chiral set of $U(1)_1$ charges for $n=6$ fermions, $\{\vec{z}_1\}$, 
 can be constructed from 4 integer parameters using the merging operation (\ref{eq:even-merger}).
We now construct a set $\{\vec{z}_2\}$ of 6 charges under a second gauge group,  $U(1)_2$, 
such that 
the combined set $[\vec{z}_1, \vec{z}_2]$ under $U(1)_1 \times U(1)_2 $ is the general solution to the anomaly equations.   
Surprisingly this is possible for any anomaly-free $\{\vec{z}_1\}$, meaning it is possible to extend any 6-fermion solution to the $U(1)_1$ anomaly equations to a solution of the $U(1)_1 \times U(1)_2$ anomaly equations.
The method we employ is to start from an arbitrary solution $\{\vec{z}_2\}$, perform a linear combination that leads to a simple set whose charges 
can be expressed in terms of the  $\{\vec{z}_1\}$ set, and then invert the linear combination
to obtain a general relation between the $\{\vec{z}_2\}$ and $\{\vec{z}_1\}$ sets. 

Using the linear combinations defined in (\ref{eq:linear-comb-gen}), we construct a set of charges with 
$\{\vec{z}_{2}\}$  replaced by a set $\{\vec{v} \, \}$  
in which  the first two charges form a vectorlike set ($v_2 = - v_1$), while $\{\vec{z}_{1}\}$ is not altered:
\begin{equation}
 [ \vec z_1 , \vec v \, ] = \left[\vec z_{1}  \; , \,    \left( z_{11}+z_{12} \right)  \vec z_2  - \left( z_{21}+z_{22} \right)  \vec z_1 \,  \right]~,
 \label{eq:linear-comb}
 \end{equation}
This is useful provided $z_{11}+z_{12} \ne 0$, which is always satisfied when $\{\vec{z}_1\}$ is chiral. Since no 4-fermion chiral set exists, the remaining charges in $\{\vec{v} \, \}$ must form two vectorlike pairs.  
Thus, we can refer to $[ \vec z_1 , \vec v \, ] $ as a chiral-vectorlike set under $U(1)_1 \times U(1)_2 $.
There are 3 possibilities for the charge pairing  in $ \{\vec{v} \, \}$:
\bear
&  
\{ v_{1}, - v_{1}, v_{3}, -v_{3}, v_{5}, -v_{5} \}  &   ~~,
\label{eq:vpairing34}
\\ [2mm]
& 
\{ u_{1}, - u_{1}, u_{3}, u_4,  -u_{3}, -u_4 \}  &   ~~,
\label{eq:vpairing35}
\\ [2mm]
&  
\{ w_{1}, - w_{1}, w_{3}, w_4,  -w_{4}, -w_3 \}  &    ~~.
\label{eq:vpairing36}
\eear
We start by analyzing the first charge pairing (\ref{eq:vpairing34}).
The mixed anomaly equations (\ref{eq:21}) and (\ref{eq:12}) now take the form
\bear
 v_3  \left(  z_{13}^2 - z_{14}^2 \right) +  v_5 \left( z_{15}^2 - z_{16}^2  \right)  & \! = \! & \! - v_1 \left( z_{11}^2 - z_{12}^2 \right)  ~~,
\label{eq:mixed12}
\\ [2mm]
 v_3^2  \left(  z_{13} +  z_{14} \right) +  v_5^2 \left( z_{15} + z_{16}  \right)  & \! = \!  & \! - v_1^2 \left( z_{11} + z_{12} \right)   ~~.
\label{eq:mixed21}
\eear
In order to solve these equations, we consider a few cases depending on the relations between the $\{\vec{z}_{1}\}$ charges.  Although a reordering of charges in $\{\vec{z}_{1}\}$ does not alter the physics, and can avoid some of the special cases discussed below, for simplicity we hold the ordering of charges fixed in our analysis.
The case where $z_{11} = z_{12} $  is discussed later in this Section.
It is convenient to use the short-hand notation 
\be
\zeta_{jj'} \equiv   \left(  z_{1j} + z_{1j'} \right)  \left(  z_{1j} - z_{1j'} \right)^2   ~~.
\ee
An identity valid for any anomaly-free set of 6 charges is  
\be
\zeta_{12} +  \zeta_{34} +  \zeta_{56} 
= - (z_{11}+z_{12}) (z_{13}+z_{14}) (z_{15}+z_{16})  ~~.
\label{eq:identity}
\ee

Multiplying Eq.~(\ref{eq:mixed21}) by $\zeta_{12} \neq 0 $ and  adding Eq.~(\ref{eq:mixed12}) squared, we obtain a quadratic equation in $v_3/v_5$:
\be
\left(  z_{13} + z_{14} \right)  \left( \zeta_{12} +  \zeta_{34} \right)  v_3 ^2     
+ 2 \left(  z_{13}^2 - z_{14}^2 \right)  \left(  z_{15}^2 - z_{16}^2 \right)   v_3 v_5
+ \left(  z_{15} + z_{16} \right)   \left( \zeta_{12} +  \zeta_{56} \right)  v_5 ^2   = 0    ~.
\label{eq:quadratic}
\ee
Let us first consider the case where $ \zeta_{12} +  \zeta_{34} \neq 0$,
which implies that the above equation in $v_3$ has two solutions.
The identity (\ref{eq:identity}) allows the simplification of the solutions to the quadratic equation.  
We label the  set $\{\vec{v} \, \}$ corresponding to each solution  
by $\{\vec{v}^{\, +} \}$ or $\{\vec{v}^{\, -} \}$.
Up to an overall normalization, their components are 
\bear
v_1^\pm & = & (z_{15}+z_{16}) \left[  -(z_{11}-z_{12})(z_{15} - z_{16}) \pm  (z_{13}^2-z_{14}^2)  \rule{0mm}{3.9mm} \right]   \label{eq:v1} \nonumber \\  [2mm] 
v_3^\pm & = & -(z_{15}+z_{16}) \left[   (z_{13}-z_{14})(z_{15} - z_{16})\pm (z_{11}^2-z_{12}^2)  \rule{0mm}{3.9mm}  \right]    \label{eq:v135} \\  [3mm] 
v_5^\pm  & = &  \zeta_{12} +  \zeta_{34}  \nonumber  
    ~~.    \nonumber 
\eear
Note that here $v_5^+ = v_5^-$, but later on this equality will not hold as we will use different normalizations for $\{\vec{v}^{\,+} \}$ or $\{\vec{v}^{\, -} \}$.

Consider now the special case $ \zeta_{12} +  \zeta_{34} = 0$, which 
combined with the identity (\ref{eq:identity}) implies $ \left(  z_{15} - z_{16} \right)^2 = -  \left(  z_{11} + z_{12} \right)   \left(  z_{13} + z_{14} \right) $.
A solution for Eq.~(\ref{eq:quadratic}) is then $v_5 = 0$, which implies $v_1 = - z_{13}^2 + z_{14}^2$ and  $v_3 = z_{11}^2 - z_{12}^2$.
This solution turns out to be the same as one of the solutions shown in Eq.~(\ref{eq:v135}) extended to the case of $ \zeta_{12} +  \zeta_{34} = 0$ (the other solution becomes trivial, $v_1 = v_3 = v_5 = 0$). 
For $v_5 \neq 0$, the quadratic equation (\ref{eq:quadratic}) becomes    
$2 \left(  z_{13}^2 - z_{14}^2 \right)  \left(  z_{15} - z_{16} \right)   v_3  = - \left( \zeta_{12} +  \zeta_{56} \right)  v_5$.
Extracting $v_1$ from Eq.~(\ref{eq:mixed12}), and removing an overall factor of $ \left(  z_{11} + z_{12} \right)  \left(  z_{13} + z_{14} \right) $,
we find the following solution for $v_1, v_3, v_5$, which we label with an upper index 0:
\bear
&&    v_1^0 = (z_{13} - z_{14} ) \left[   ( z_{11} -  z_{12}  )^2 + ( z_{13} +  z_{14}  ) ( z_{15} +  z_{16}  ) \right]  ~~,
 \nonumber \\  [2mm] 
&&    v_3^0 = (z_{11} - z_{12} ) \left[   ( z_{13} -  z_{14}  )^2 + ( z_{11} +  z_{12}  ) ( z_{15} +  z_{16}  ) \right] ~~,
\label{eq:zeta0}
 \\  [2mm] 
&&    v_5^0 = 2   ( z_{11} - z_{12}) ( z_{13} - z_{14} )  ( z_{15} - z_{16} ) ~~.   \nonumber  
\eear
This solution cannot be obtained from the ones given in (\ref{eq:v135}), even after allowing an arbitrary change in the overall normalization.

In the special case $z_{11} = z_{12} $, for any value of  $ \zeta_{12} +  \zeta_{34}$, 
Eqs.~(\ref{eq:mixed12}) and   (\ref{eq:mixed21})  have the solutions $v_1 = \pm  ( z_{13} +  z_{14}  ) ( z_{15} +  z_{16}  ) $, 
$v_3 = - z_{15}^2 + z_{16}^2$, $v_5 =  z_{13}^2 -  z_{14}^2$. Note that these can be obtained from the solutions (\ref{eq:v135})
when extended to the case $z_{11} = z_{12} $. 
We conclude that there are only two nontrivial solutions for $v_1,v_3,v_5$, namely the 
two solutions shown in Eqs.~(\ref{eq:v135}),  or one of those solutions and Eq.~(\ref{eq:zeta0}) in the special case $ \zeta_{12} +  \zeta_{34} = 0$.  
 
From Eq.~(\ref{eq:linear-comb}) it  follows that for any anomaly-free set of 6 charges under $U(1)_1$, $\{\vec{z}_1 \}$, the 
sets of $U(1)_2$  charges which satisfy all anomaly equations are
\be
 \vec z_2^{\; \pm, 0}  =  \frac{\kappa}{z_{11}+z_{12}} \, \vec v^{\, \pm, 0} +   \frac{z_{21}+z_{22} }{z_{11}+z_{12} } \,  \vec z_1   ~~,
\label{eq:z2shift}
\ee
where $\kappa $ is a parameter that allows for changes in the normalizations of $\vec v^{\, \pm, 0}$ displayed in Eqs.~(\ref{eq:v135}) and (\ref{eq:zeta0}).  By rescaling $\vec v^{\, \pm, 0}$ and $\vec z_1$ their coefficients in (\ref{eq:z2shift}) can always be made integer.  Furthermore, from Eq.~(\ref{eq:linear-comb-gen}) any integer choice for the coefficients will lead to a solution for $\vec z_2^{\; \pm, 0}$.  Thus, the most general sets of $U(1)_2$ charges which satisfy all anomaly equations are
\be
 \vec z_2^{\; \pm, 0}  = \kappa \, \vec v^{\, \pm, 0} + \lambda \,  \vec z_1   ~~,
\label{eq:z2shiftgeneral}
\ee
with $\kappa$ and $\lambda$ integers and the most general solutions to the system of four anomaly equations 
(for two mixed $U(1)_1 \times U(1)_2$ anomalies, and two  $U(1)_2$  anomalies)  
are the following sets:  
\bear
[\vec{z}_1, \vec{z}_2^{\; \pm, 0} ] =   
\begin{bmatrix}
\begin{array}{c|c}
z_{11} & \,   \kappa \,   v_1^{\pm, 0} + \lambda\, z_{11}\\
z_{12} & -\,   \kappa \,  v_1^{\pm, 0}  + \lambda\, z_{12}\\
z_{13} & \,  \kappa \,  v_3^{\pm, 0}  + \lambda\, z_{13}\\
z_{14} & -\,  \kappa \,  v_3^{\pm, 0}  + \lambda\, z_{14}\\
z_{15} & \, \kappa \,  v_5^{\pm, 0}  + \lambda\, z_{15}\\
z_{16} & -\,  \kappa \,  v_5^{\pm, 0} + \lambda\, z_{16}\\
\end{array}
\end{bmatrix},\\ \nonumber
\label{eq:U(1)U(1)}
\eear
where the $\pm, 0$ upper indices are correlated, \ie , there is a solution with index $+$, another solution
with index  $-$, and a special solution with index 0 that applies only for a set $  \vec z_1  $
that satisfies $ \zeta_{12} +  \zeta_{34} = 0$, in which case only one of $v^{\, \pm}$ is nontrivial.
Without loss of generality, given the freedom of redefining the $U(1)_2$  gauge coupling, we take the parameters  $\kappa$ and $\lambda$ as integers.
One should keep in mind  that in order to obtain a coprime set
the six integers that represent the $U(1)_2$ charges 
shown above may need to be divided by their greatest common divisor.
Note that the $[ \vec z_1 , \vec z_2 \, ] $ sets are typically chiral under each of the $U(1)$'s; we refer to such sets as 
doubly-chiral under $U(1)_1 \times U(1)_2 $.

We now turn to the second charge pairing in the set $\{\vec{v} \, \}$, which is given in Eq.~(\ref{eq:vpairing35}).
Following the same steps as for the (\ref{eq:vpairing34}) pairing, we find that (up to an overall normalization) 
the components $u_1, u_3, u_4$ of $\{\vec{v} \, \}$ are related to $v_1, v_3, v_5$ given in (\ref{eq:v135}) by an interchange of $z_{14}$ and   $z_{15}$.

In the special case where the $\{ \vec z_1 \} $ set satisfies $\zeta_{12} +  \zeta_{35}  = 0$, there is an additional solution
obtained from (\ref{eq:zeta0}) by interchanging $z_{14}$ and $z_{15}$.
Similarly, the solutions for the (\ref{eq:vpairing36}) pairing  are obtained 
from Eqs.~(\ref{eq:v135}) and (\ref{eq:zeta0}) by replacing $v_1$, $v_3$, $v_5$
with  $w_1$, $w_3$, $-w_4$, and 
by interchanging $z_{14}$ and $z_{16}$.
Thus, in addition to the two familes of $U(1)\times U(1)$ sets shown in (\ref{eq:U(1)U(1)}), there are four more families of anomaly-free sets:
\begin{equation}
\begin{bmatrix}
\begin{array}{c|c}
z_{11} & \,   \kappa \,   u_1^{\pm, 0} + \lambda\, z_{11}\\
z_{12} & -\,   \kappa \,  u_1^{\pm, 0}  + \lambda\, z_{12}\\
z_{13} & \,  \kappa \,  u_3^{\pm, 0}  + \lambda\, z_{13}\\
z_{14} & \,  \kappa \,  u_4^{\pm, 0}  + \lambda\, z_{14}\\
z_{15} & - \, \kappa \,  u_3^{\pm, 0}  + \lambda\, z_{15}\\
z_{16} & -\,  \kappa \,  u_4^{\pm, 0} + \lambda\, z_{16}\\
\end{array}
\end{bmatrix}   \;\;\; ,  \;\;\; \;\;\; 
\begin{bmatrix}
\begin{array}{c|c}
z_{11} & \,   \kappa \,   w_1^{\pm, 0} + \lambda\, z_{11}\\
z_{12} & -\,   \kappa \,  w_1^{\pm, 0}  + \lambda\, z_{12}\\
z_{13} & \,  \kappa \,  w_3^{\pm, 0}  + \lambda\, z_{13}\\
z_{14} & \,  \kappa \,  w_4^{\pm, 0}  + \lambda\, z_{14}\\
z_{15} & -\, \kappa \,  w_4^{\pm, 0}  + \lambda\, z_{15}\\
z_{16} & -\,  \kappa \,  w_3^{\pm, 0} + \lambda\, z_{16}\\
\end{array}
\end{bmatrix}  ~~.
\label{eq:U(1)U(1)uw}
\end{equation}
Thus, the general solution is defined by 6 integer parameters, 4 integers necessary to specify $\{\vec{z}_{1}\}$, $\kappa$, and $\lambda$, and a discrete choice from six solutions, two for each choice of charge pairing in $ \{\vec{v} \, \}$.  We will refer to these discrete choices as ``branches".
\bigskip

\subsection{Parametrization of the $U(1)\times U(1)$ solutions}
\label{sec:kkll}

We now construct the parametrization of the chiral-vectorlike and doubly-chiral solutions for 6 fermions charged under 
$U(1)_1\times U(1)_2$.  The general  set of charges $\{\vec{z}_1\}$, which is a solution
to the  $U(1)_1$ anomaly equations, depends on 4  integer parameters, $k_{1,2}, \ell_{1,2}$,  as follows from Eq.~(\ref{eq:even-merger})
and the detailed discussion for 6-charge sets given in \cite{Costa:2019zzy}.
Since the $z_{1j}$ are quartic polynomials in the $k_{1,2}, \ell_{1,2}$, one would expect the $\vec z_2^{\; \pm}$ charges  to be collectively 12$^\mathrm{th}$ order in the $k_{1,2}$ and $\ell_{1,2}$ parameters but various common polynomial factors arise allowing simplification.  
The simplification is maximal when the last three charges of the solution generated from the merger (\ref{eq:even-merger}) are reordered.
 More precisely, we take 
 \be
 \{ \vec{z}_1\, \} =  \{  \ell_1, k_1, k_2, - k_2,  - \ell_1, -k_1  \}  \oplus  \{  0, 0, \ell_1, - \ell_2,  \ell_2, - \ell_1 \}~~,
 \ee
 which gives the six $U(1)_1$ charges as the following quartic polynomials: 
\bear
&&  \hspace*{-0.3cm} 
z_{11} = \ell_1  \left(  \ell_1^2 ( k_2 - k_1 ) - \ell_2^2 (\ell_1 + k_2 )   \rule{0mm}{4mm} \right) 
\nonumber \\ [1.2mm]
&&  \hspace*{-0.3cm} 
z_{12} =  k_1 \left(  \ell_1^2  (k_2 - k_1) - \ell_2^2 (\ell_1+k_2 )    \rule{0mm}{4mm} \right)
\nonumber \\ [2mm]
&&  \hspace*{-0.3cm} 
z_{13} = \ell_1^2 k_1 ( k_1 - k_2) -  \ell_2 (\ell_1 + k_2) (\ell_1^2 -  \ell_1 k_2 + k_2 \ell_2)
\nonumber \\ [-2.mm]
\label{eq:6set}
\\ [-3mm]
&&  \hspace*{-0.3cm} 
z_{14} =  \ell_1 \left(    \ell_2 ^2(\ell_1 + k_2 )   +  (k_2 -k_1) ( k_2 \ell_2 - \ell_1 k_2 + k_1 \ell_2 )   \rule{0mm}{4mm} \right)
\nonumber \\ [2mm]
&&  \hspace*{-0.3cm} 
z_{15} = \ell_2^2 k_2 (\ell_1 + k_2 ) -  \ell_1 (k_2 - k_1 ) (\ell_1^2 + k_1 \ell_2 + k_2 \ell_2)
\nonumber \\ [2mm]
&&  \hspace*{-0.3cm} 
z_{16} = \ell_1^2 k_2 (k_2 - k_1 ) +  \ell_2 ( \ell_1 + k_2 ) (\ell_1^2 - \ell_1 k_2 + k_1  \ell_2 )~~.
\nonumber 
\eear
Inserting this parametrization in Eq.~(\ref{eq:v135}) we obtain the components of the vectorlike set $\{ \vec{v}^{\, +}\}$ after removing the greatest common divisor (which is an 8th degree polynomial in the integer parameters):
\begin{eqnarray}   \hspace*{-1.6cm}
&& v_{1}^{+} = \ell_2^2(k_2+\ell_1)^2+\ell_1\ell_2(k_2+\ell_1)(\ell_1 \! - \! 2k_2 \! - \! k_1)  -  \ell_1^4 + \ell_1^2 k_2^2 - \ell_1^2  ( k_1 \!-\!  k_2 ) (k_1 \!-\!  \ell_1)  ~,
 \nonumber \\ [1.5mm]  \hspace*{-1.6cm}
&& v_{3}^{+} = \ell_2^2(k_2+\ell_1)^2 + \ell_1\ell_2(k_2+\ell_1)(k_1+\ell_1) + \ell_1^4 - \ell_1^2 k_1^2 +\ell_1^2 (k_1 \!-\!  k_2)(k_2+\ell_1)  ~,
\label{eq:vset} \\  [1.5mm]    \hspace*{-1.6cm}
&& v_{5}^{+} = -\ell_2^2(k_2+\ell_1)^2+\ell_1\ell_2(k_2+\ell_1)(k_1+\ell_1) +  \ell_1^4 - \ell_1^2 k_1^2  -  \ell_1^2 (k_1 \!-\!  k_2)(k_2+\ell_1) 
~~. \nonumber 
\end{eqnarray}
The set  $[\vec{z}_1, \vec{z}_2^{\; +} ] $ shown in 
Eq.~(\ref{eq:U(1)U(1)}) is a solution to the $U(1)_1\times U(1)_2$ anomaly equations for 6 fermions
provided  the  components of  $\vec{z}_1$ are parametrized as in Eq.~(\ref{eq:6set})
while the components of $\{\vec{v}^{\, +} \}=\{ v_{1}^+, - v_{1}^+, v_{3}^+, -v_{3}^+, v_{5}^+, -v_{5}^+ \} $
are parametrized as in Eq.~(\ref{eq:vset}).
Altogether, this solution depends on 6 integer parameters: $k_1, k_2, \ell_1, \ell_2, \kappa, \lambda$.

Instead of using the $\{\vec{v}^{\, +} \}$ vectorlike set, one could construct a different solution with the $\{\vec{v}^{\, -} \}$ set, whose  
 components are similarly given by 
\begin{eqnarray}    \hspace*{-1.4cm}
&& v_{1}^{-} =  -\ell_1^2 (k_1 \!-\! k_2)^2 - \ell_1\ell_2(k_1 \!-\!  k_2)(k_1 \! + \! 2k_2 \! - \! \ell_1 \! +  \! \ell_2) +  \ell_2^2  (k_1^2 - k_1 k_2 -  k_2^2 + \ell_1^2) 
 ~,
 \nonumber \\ [1.5mm]     \hspace*{-1.4cm}
&& v_{3}^{-} =  - \ell_1^2(k_1 \!-\!  k_2)^2 - \ell_1\ell_2(k_1 \!-\! k_2)(k_1+\ell_1 +\ell_2 )  - \ell_2^2 \left(k_1^2 + k_1 k_2 - k_2^2 -\ell_1^2 \right)  
 ~,
\label{eq:vsetm}  \\  [1.5mm]   \hspace*{-1.4cm}
&&  v_{5}^{-} =   \ell_1^2(k_1 \!-\!  k_2)^2 - \ell_1\ell_2(k_1 \!-\! k_2)(k_1+\ell_1 - \ell_2)   - \ell_2^2 \left(k_1^2 - k_1 k_2 + k_2^2-\ell_1^2 \right)  
 ~~.  \nonumber
\end{eqnarray}
In the case where $\zeta_{12} +  \zeta_{34} = 0$,  the $[\vec{z}_1, \vec{z}_2^{\; 0} ]$  solution from Eq.~(\ref{eq:U(1)U(1)})
is also a set of six polynomials in the integer parameters.
Likewise, the families of anomaly-free sets with different pairings, shown in Eq.~(\ref{eq:U(1)U(1)uw}),
can be written in terms of polynomials in  $k_1, k_2, \ell_1, \ell_2, \kappa, \lambda$.

We now present some numerical examples of solutions for $U(1)_1\times U(1)_2$ with 6 fermions.  We start from a primitive solution to the single $U(1)$ anomaly equations, $\{\vec{z}_1\}$ as given in (\ref{eq:6set}),  and generate the families of $U(1)_1\times U(1)_2$ anomaly-free  sets 
$[\vec{z}_1,\, \kappa \vec{v} + \lambda  \vec{z}_1]$ as in Eqs.~(\ref{eq:U(1)U(1)})
and (\ref{eq:U(1)U(1)uw}).
The first two chiral solutions \cite{Costa:2019zzy} with the smallest maximal charge are, in canonical form, $\{\vec{z}_1\}=\{5,\, -4, \, -4,\, 1,\, 1,\, 1\}$ and $\{\vec{z}_1\}=\{ 6, -5, -5, 3, 2, -1\}$. These  
are generated upon a charge reordering from the mergers with $(k_1,\,k_2,\,\ell_1,\,\ell_2) = (1,\,-2,\,1,\,2)$ and  $(2, 0, 1, -1)$, respectively.  
Since any multiple of the vectorlike set $\{\vec{v} \}$ is also a solution to the anomaly equations, we choose to present its charges as a coprime set with the first entry (corresponding to the fermion with largest $U(1)_1$ charge) positive. 
Note that the reordering that puts  $\{\vec{z}_1\}$ in canonical form also changes the  ordering within 
$\{\vec{v} \, \}$.
In the case of $\{\vec{z}_1\}=\{5,\, -4, \, -4,\, 1,\, 1,\, 1\}$, due to the degeneracies among its charges,
only two of the six vectorlike branches are distinct, and the only $U(1)_1\times U(1)_2$ anomaly-free sets are 
\begin{equation}
\begin{bmatrix}
\begin{array}{c|c}
5 & 5\kappa + 5\lambda \\
-4 & -5\kappa - 4 \lambda \\
-4 & -3\kappa - 4\lambda \\
1 &  3 \kappa + \lambda \\
1 & \kappa + \lambda \\
1 & - \kappa + \lambda
\end{array}
\end{bmatrix}  
\;\;\;\; , \;\;\;\;  \;  
\begin{bmatrix}
\begin{array}{c|c}
5 &   5\lambda \\
-4 & \kappa - 4 \lambda \\
-4 & - \kappa - 4\lambda \\
1 &  2 \kappa + \lambda \\
1 & - 2 \kappa + \lambda \\
1 &   \lambda
\end{array}
\end{bmatrix}   ~~.
\label{eq:544111}
\end{equation}
For $\{\vec{z}_1\}=\{ 6, -5, -5, 3, 2, -1\}$ there are 3 distinct vectorlike branches: $\{1, 3, -3, -1, -9, 9\}$, 
$\{9, 1, -11, -9, -1, 11\}$ and  
$\{3, 1, -5, -3, 5, -1\}$. Each of these vectorlike sets generates a 2-parameter family of anomaly-free  sets,
$[\vec{z}_1,\, \kappa \vec{v} + \lambda  \vec{z}_1]$. 
For other choices of $\{ \vec{z}_1 \}$, such as $\{11, -10, -8, 5, 4, -2\}$, all 6 vectorlike branches are distinct.

\subsection{Example of a chiral dark sector}
\label{sec:dm}

The $U(1)_1 \times U(1)_2$ chiral sets obtained here may have interesting applications in a variety of hidden-sector models,
related to right-handed neutrinos, dark matter, and other phenomena.
To give an example, we take the second set of (\ref{eq:544111}), fix $\lambda = 0$ and $\kappa = 1$,
and introduce a scalar $\phi$ carrying charges $(-2,0)$ under $U(1)_1\times U(1)_2$. 

The field content of this model is shown in Table~\ref{table:DM}.
The gauge charges allow the Weyl fermions $\psi_i$, $i = 1,... , 6$, to have 
the following Lagrangian interactions with the scalar:
\be
- y_{45} \, \phi \, \psi_4 \psi_5 
-  \frac{c_{16} }{M_*^2} \,  \phi^3 \, \psi_1 \psi_6 
-  \frac{c_{23} }{M_*^3}  \left( \phi^\dagger \right)^4 \psi_2 \psi_3 
+ {\rm H.c.} 
\ee
The first term is renormalizable, and its dimensionless coefficient, the Yukawa coupling $y_{45} > 0$, 
is presumably of order one.
The last two terms are effective interactions generated at the scale $M_*$ where the two $U(1)$ groups 
are embedded in some non-Abelian structure. We will not provide an explicit renormalizable origin of these two terms,
but they may be generated at tree level, or at loop level, by 
various heavy particles which are integrated out.
The two dimensionless coefficients, $c_{16}, c_{23} > 0$, may thus be of order one or much smaller than that.
 
A scalar potential with a minimum at $\langle \phi \rangle > 0 $ leads to three Dirac masses.
The 4-component fermion formed of $ \psi_2$ and $ \psi_3$ has  a mass $m_{23} = c_{23} \langle \phi \rangle^4/M_*^3$.
Similarly, the Dirac  fermion formed of $ \psi_1$ and $ \psi_6$ has  a mass $m_{16} = c_{16} \langle \phi \rangle^3/M_*^2$,
while the remaining mass is $m_{45} = y_{45} \langle \phi \rangle$.
Parametrically, we expect large hierarchies between these masses: 
\be
m_{23} \ll m_{16} \ll m_{45}  ~~.
\ee

\begin{table}[t!]
\begin{center}
\renewcommand{\arraystretch}{1.5}
\begin{tabular}{|c||c|c|c|}\hline  
Field  &  Spin  &   $U(1)_1$   &   $U(1)_2$ 
\\ \hline \hline
    $\psi_1$   &  1/2  &  5  &  0
\\ [1mm]    \hline
    $\psi_2$   &  1/2  &  $-4$  &  1
\\ [1mm]    \hline
    $\psi_3$   &  1/2  &  $-4$  &  $-1$
\\ [1mm]    \hline
    $\psi_4$   &  1/2  &  1  &  $2$
\\ [1mm]    \hline
    $\psi_5$   &  1/2  &  1  &  $-2$
\\ [1mm]    \hline
    $\psi_6$   &  1/2  &  1  &  0
\\ [1mm]    \hline\hline
    $\phi$      &  0       &  $-2$  &  0
\\ [1mm]    \hline
  \end{tabular}
\caption{A dark-sector model with $U(1)_1 \times U(1)_2$ gauge symmetry, 
6 left-handed fermions, and a scalar. The fermions form three stable Dirac states with hierarchical masses.}
\label{table:DM}
\end{center}
\end{table}

These 3 Dirac fermions are stable due to a global $U(1)^3$ symmetry. Thus, the model described here is 
a good candidate for a dark sector, with three  dark matter components.
The gauge boson associated with $U(1)_2$ is a massless dark photon, while
the one associated with $U(1)_1$ is a massive ``dark $Z$" boson.
Only the lightest  and the heaviest fermions interact with the  dark photon, while all fermions interact with the dark $Z$. 

It is interesting that even one of the simplest chiral fermion set leads to such a complex dark sector. 
If the dark matter particles are described by this model, then there are opportunities for discovery 
in various experiments probing very different scales.

\bigskip

\subsection{Doubly-vectorlike solutions}
\label{sec:doublyvec} 

Let us now construct a solution where both $\{\vec{ z }_{1}\}$ and $\{\vec{ z }_{2}\}$  are vectorlike sets under their respective $U(1)$, but the $U(1)_1 \times U(1)_2$ set is chiral.  
By reordering the fermions, we can always put $\{\vec{ z }_{1}\}$ in the following form:
$\{\vec{z}_1\} = \{ j_{1}, -j_{1}, j_{2}, -j_{2}, j_{3}, -j_{3}\}$,   
where  $j_1,j_2,j_3$ are 
integer parameters with $j_1 \ge j_2 \ge j_3 \ge 0$. 
After a redefinition of the $U(1)_1$ gauge coupling, 
the GCD of $j_1,j_2,j_3$ is taken to be 1.

\begin{table}[t!]
\begin{center}
\renewcommand{\arraystretch}{1.3}
\vspace*{0.4cm}
\begin{tabular}{c|c|c|c|c|c|c|c}\hline  
\hspace*{-0.4cm} $[\vec{z}_1, \vec{z}_2 ] $ & \multicolumn{7}{c}{ the other 7 solutions for the $\vec{z}_2$ set }
\\ [1mm] \hline   \\ [-4mm] 
\hspace*{-0.2cm}
$  \!\!  \left[   \!\!
\begin{array}{c|c}
 j_{1} & \,  j'_{1} \\
-   j_{1} & \, j'_{2} \\
 j_2 & \, - j'_1  \\
-   j_2 &  j'_3 \\
 j_3 &  -  j'_{2} \\
-    j_3  & -  j'_3 \\
\end{array}    \! \right] $  
& \multicolumn{7}{c}{
$ \left.  \begin{array}{c}
 j'_{1} \\
 j'_{2} \\
 j'_3 \\
- j'_{2} \\
- j'_3 \\
- j'_1  \\
\end{array} \! \right] $    \hspace{0.35cm}
$ \left.  \begin{array}{c}
 j'_1 \\
 j'_2 \\
 j'_3  \\
- j'_1 \\
- j'_3 \\
- j'_2 \\
\end{array} \! \right] $   \hspace{0.4cm}
$ \left.  \begin{array}{c}
 j'_1 \\
 j'_2 \\
 - j'_2  \\
 j'_3 \\
- j'_1 \\
- j'_3 \\
\end{array}\! \right] $   \hspace{0.4cm}
$ \left.  \begin{array}{c}
 j'_{1} \\
 j'_{2} \\
- j'_1  \\
 j'_3 \\
 - j'_3 \\
 - j'_2 \\
\end{array} \! \right] $    \hspace{0.45cm}
$ \left.  \begin{array}{c}
 j'_{1} \\
 j'_{2} \\
 j'_3  \\
- j'_2 \\
- j'_1 \\
- j'_3 \\
\end{array} \! \right] $     \hspace{0.45cm}
$ \left.  \begin{array}{c}
 j'_{1} \\
 j'_{2} \\
 j'_3  \\
 -  j'_1 \\
 -  j'_{2} \\
 -  j'_3 \\
\end{array} \! \right] $     \hspace{0.45cm}
$ \left.  \begin{array}{c}
 j'_1 \\
 j'_2 \\
 - j'_2  \\
 j'_3 \\
- j'_3 \\
- j'_1 \\
\end{array} \! \right] $  \hspace{-0.35cm}    } 
\\  [-5mm]    \\     \hline 
$ j'_1,j'_2,j'_3$    &  &   $ \!  j_1   \!  \to  \!  -j_1 $   &  $j_1   \!  \to \! -j_1 $  &  $j_3 \!  \to  \!  -j_3 $  &   $j_3  \!  \to  \!  -j_3 $   &  \ $j_2  \!  \to  \!  -j_2 $   &   $j_2  \!  \to  \!  -j_2 \! \! $
\\   \cline{2-8}
$\!\! $ from (\ref{eq:quadratic-pol})   \      &  $ \!   j'_1 \leftrightarrow j'_2$  & &   $j'_1 \leftrightarrow j'_2$    &   &   $j'_1 \leftrightarrow j'_2$   &  &   $j'_1 \leftrightarrow j'_2\! \!  $   
\\ [1mm]  \hline
\end{tabular}
 \vspace{0.1cm}
\caption{All doubly-vectorlike solutions to the $U(1)_1\times U(1)_2$ anomaly equations for 6 chiral fermions.  The  $[\vec{z}_1, \vec{z}_2 ] $ solution 
shown in the first column  depends only on the $j_1,j_2,j_3$ integer parameters, with the $\vec{z}_2$ set of $U(1)_2$ charges written in terms of 
$j'_1,j'_2,j'_3$, which are the quadratic polynomials (\ref{eq:quadratic-pol}) in $j_1,j_2,j_3$. The other 7 columns represent all remaining
solutions for $\vec{z}_2$, with $j'_1,j'_2,j'_3$ for each of them obtained from  (\ref{eq:quadratic-pol}) by flipping the sign for one of 
the $j$ parameters (specified in the second-to-last row), or by interchanging $j'_1$ and $j'_2$, or in three cases by performing both transformations.
 }
\label{table:2}
\end{center}
\end{table}

The second vectorlike set, $\{\vec{ z }_{2}\}$, will also depend on three integer parameters, $j'_1,j'_2,j'_3$.
The first pair of $\{\vec{ z }_{2}\}$ cannot be vectorlike (because we seek the cases 
where the $[\vec{ z }_{1},\vec{ z }_{2}]$ set is chiral).
Thus, the first two charges in $\{\vec{ z }_{2}\}$  can be taken to be $j'_1$ and $j'_2$. 
There are four possible choices for the next pair of entries in $\{\vec{ z }_{2}\}$: $(j'_3, - j'_1)$,  $(j'_3, - j'_2)$, $(-j'_1, j'_3)$, or
$(- j'_2, j'_3)$.  For each of these four choices there are two possible choices for the last two entries 
in $\{\vec{ z }_{2}\}$. As a result, there are up to 8 distinct sets of charges which are anomaly-free and doubly-vectorlike under $U(1)_1 \times U(1)_2$, as shown in Table \ref{table:2}.
Following the terminology of Section \ref{sec:chi-chi}, these sets are branches of the general solution to the anomaly equations.

For 
the first branch of this type, the anomaly equations determine $j'_1, j'_2,  j'_3$ in terms of $j_1, j_2,  j_3$ as follows:
\begin{eqnarray}
&& j'_1 =  j_1^2   + j_2^2 - j_3^2 - j_1 j_2 - j_2 j_3  + j_1 j_3 ~~, \nonumber \\
&& j'_2 = - \left(  j_1^2 - j_2^2 + j_3^2 - j_1 j_2 - j_2 j_3 + j_1 j_3 \right)  ~~,   \label{eq:quadratic-pol} \\
&& j'_3 = - j_1^2 + j_2^2 + j_3^2  - j_1 j_2 - j_2 j_3  + j_1  j_3 ~~. \nonumber 
\end{eqnarray}
For each of the other 7 arrangements of the second set $\{\vec{z}_2\}$, the expressions for  $j'_1,j'_2,j'_3$
in terms of $j_1,j_2,j_3$ are obtained from the above equations by flipping the sign of one of the 
$j_1,j_2,j_3$ parameters, or by the interchange $j'_1 \leftrightarrow j'_2$.
In three of these cases, as indicated in Table \ref{table:2}, both the sign flip and the interchange need 
to be performed on Eqs.~(\ref{eq:quadratic-pol}) in order to obtain the correct $j'_1,j'_2,j'_3$ dependence on $j_1,j_2,j_3$.

If $j_1 = j_2$, then Eqs.~(\ref{eq:quadratic-pol}) give  $j'_1 =  j'_2  = - j'_3 $, which makes
 the $[\vec{ z }_{1},\vec{ z }_{2}]$ set vectorlike. Likewise,  
if $j_2= j_3$, then $- j'_1 =  j'_2  =  j'_3 $, leading to a vectorlike set. It is thus sufficient to impose
\be
j_1 > j_2 > j_3 \ge 0  ~~. 
\ee

Out of the 8 possible solutions for $\vec{z}_{2}$ shown in Table~\ref{table:2}, only 3 are linearly independent.   
Nevertheless, when the masses of the two $U(1)_1 \times U(1)_2$ gauge bosons are different,
each of those 8 solutions generically leads to different observables,
as the couplings between the 
gauge bosons and fermions are determined by the charges.
As a numerical example, $j_1 = 3$, $j_2 = 2$, $j_3  = 1$ gives the following 8 solutions: \\
\bear    \hspace{-0.7cm}
 \!\!  \left[   \!\!
\begin{array}{c|c}
 3 & \,  7 \\
- 3 & \, -1 \\
 2 & \, - 7  \\
- 2 & -9 \\
 1 &  1 \\
- 1  & 9 \\
\end{array}    \! \right]     \hspace{0.4cm}
 \left.  \begin{array}{c}
 -1 \\
 7 \\
 -9 \\
- 7 \\
9 \\
1  \\
\end{array} \! \right]  \hspace{0.3cm}
\left.  \begin{array}{c}
 13 \\
 -7  \\
 -3  \\
- 13 \\
  3 \\
  7 \\
\end{array} \! \right]   \hspace{0.3cm}
 \left.  \begin{array}{c}
 -7  \\
 13 \\
 - 13  \\
 - 3 \\
 7 \\
 3 \\
\end{array}\! \right]  \hspace{0.3cm}
 \left.  \begin{array}{c}
 5 \\
 1 \\
- 5  \\
- 11 \\
 11 \\
 - 1 \\
\end{array} \! \right]   \hspace{0.3cm}
 \left.  \begin{array}{c}
 1 \\
 5 \\
- 11  \\
- 5 \\
 -1 \\
 11 \\
\end{array} \! \right]     \hspace{0.3cm}
 \left.  \begin{array}{c}
 23 \\
 -17 \\
  7  \\
 - 23 \\
 17 \\
 - 7 \\
\end{array} \! \right]      \hspace{0.3cm}
 \left.  \begin{array}{c}
 -17 \\
  23  \\
 - 23  \\
 7  \\
- 7 \\
 17 \\
\end{array} \! \right]  ~, 
\\ \nonumber
\eear
 where we show $[\vec{z}_1, \vec{z}_2 ] $ for the first solution,  
and only $\{\vec{z}_2\}$ for the other 7 solutions (as $\vec{z}_1$ remains unchanged).

When $ j_3 = 0$ only 4 of the 8 solutions are distinct (for example, the third and seventh solutions 
shown in Table \ref{table:2} become identical).
In the particular case $j_1 = 2$, $j_2 = 1$, $j_3  = 0$, the four solutions are  \\
\bear    \hspace{-0.1cm}
 \!\!  \left[   \!\!
\begin{array}{c|c}
 2 & \,  3 \\
- 2 & \, - 1 \\
 1 & \, - 3  \\
- 1 & - 5 \\
 0 &  5 \\
 0  & 1 \\
\end{array}    \! \right]     \hspace{0.8cm}
 \left.  \begin{array}{c}
 -1  \\
 3 \\
 -5  \\
 - 3 \\
 5 \\
 1 \\
\end{array}\! \right]     \hspace{0.7cm}
 \left.  \begin{array}{c}
   7 \\
 -5 \\
- 1  \\
- 7 \\
 5 \\
 1 \\
\end{array} \! \right]  \hspace{0.7cm}
\left.  \begin{array}{c}
 -5  \\
 7  \\
 -7  \\
  -1 \\
  5 \\
  1 \\
\end{array} \! \right] 
 ~~~.  \\ \nonumber
\eear
Note that doubly-vectorlike and chiral-vectorlike solutions can be related by taking linear combinations as in Eq.~(\ref{eq:linear-comb-gen}).  For instance, a reordering of the second doubly-vectorlike solution above, $\{\vec{z}_1^{\,\,VV}\}=\{0,-1,1,2,0,-2\}$, $\{\vec{z}_2^{\,\,VV}\}=\{5,-5,-3,3,1,-1\}$, can be converted to the first chiral-vectorlike solution in Eq.~(\ref{eq:544111}) with $\kappa=1$ and $\lambda=0$, since $\{\vec{z}_1^{\,\,CV}\}=\{\vec{z}_2^{\,\,VV}\}-\{\vec{z}_1^{\,\,VV}\} = \{5,-4,-4,1,1,1\}$ and $\{\vec{z}_2^{\,\,CV}\}=\{\vec{z}_2^{\,\,VV}\}$.

Although we focus on theories with the smallest chiral sets, it is instructive to comment on larger sets of fermions charged under $U(1)\times U(1)$.  
For a general doubly-vectorlike solution $[ \vec z_1 , \vec z_2]$ with $n=2p$ fermions,
where $p \in \mathbb{Z}$, the number $b(p)$ of possible arrangements of two sets of vectorlike charges such that the solution is chiral satisfies a recurrence relation:
\be
b(p) = 2(p-1)\left[b(p-1)+b(p-2)\right]\quad \mathrm{with}\quad b(1)=0,\, b(2)=2~.
\label{eq:bp}
\ee
This expression for $b(p)$ reproduces the previously discussed result for $n=6$, \ie, $b(3)=8$, and can be proven inductively as follows.
Order the first set in the form $\{\vec{z}_1\}=\{ j_{1}, -j_{1}, j_{2}, -j_{2},\ldots j_{p+1}, -j_{p+1}\}$, which can be done without loss of generality.
Then take one of the $b(p+1)$ chiral branches for $n=2(p+1)$ fermions and interchange two charges in $\{\vec{z}_2\}$ such that $[ \vec z_1 , \vec z_2]$ is no longer chiral.  
We can always order the fermions such that the last two entries in $[ \vec z_1 , \vec z_2]$ are vectorlike. Since we started from a chiral solution, this may happen in two ways: the swap turns $\{\vec{z}_2\}$ into one vectorlike pair and a chiral solution of length $2p$, or into two vectorlike pairs and a chiral solution of length $2(p-1)$.  We can reverse this procedure and construct all $b(p+1)$ chiral branches by starting from a composite solution containing a chiral set and either one or two vectorlike pairs and then carry out one swap.  In the first case there are $2p$ ways to swap one entry of the vectorlike pair with any of the charges from the chiral set, and $b(p)$ chiral sets.  In the second case there are $p$ possibilities for position of the second vectorlike pair, $b(p-1)$ chiral sets, and 2 ways to carry out a swap within the two vectorlike pairs to make the whole solution chiral.  Thus, $b(p+1) = 2p\, b(p) + 2p\, b(p-1)$, which completes the proof of Eq.~(\ref{eq:bp}). 

This implies that the number of branches $b(p)$ for doubly-vectorlike solutions grows quickly with the number $n=2p$ of fermions, \eg, $b(4)=60$ and $b(5)=544$.  
The sequence of integers generated by the recurrence relation (\ref{eq:bp}) arises in various combinatorics problems \cite{oeis}.
For $p \gg 1$, we find the following approximate behavior: $b(p)\approx c\, (2p-1)!!$ with $c \approx 0.6$.

To summarize the main result of this Section, we have identified the general solution for the 
charges of 6 fermions which form an anomaly-free chiral set under the $U(1)\times U(1)$ gauge group.
The general doubly-vectorlike solution depends only on the three integer parameters $j_1,j_2,j_3$, 
and on a discrete parameter that labels up to 8 branches shown in Table~\ref{table:2}.
The general doubly-chiral solution  
depends on six integer parameters, and has 6 branches 
given in (\ref{eq:U(1)U(1)}) and (\ref{eq:U(1)U(1)uw}). Any chiral-vectorlike solution can be obtained from the 
doubly-chiral solution by setting  $\lambda=0$. 

\bigskip \bigskip \bigskip

\section{Three or more $U(1)$ gauge groups}
 \setcounter{equation}{0}\label{sec:3U1}

We now turn to a theory with $n$ Weyl fermions charged under a $U(1)^m$ gauge group where $m \ge 3$.
The number of anomaly equations,  ${\cal N}_{\rm eq}$,
is given in Eq.~(\ref{eq:neq}).
In the case $m=3$,  \ie, when the gauge group is $U(1)_1\times U(1)_2 \times U(1)_3$,
there are 13 anomaly equations: a cubic equation (\ref{eq:cubic})  and a linear equation (\ref{eq:linear})  for each $U(1)$,  
two mixed equations, (\ref{eq:21}) and (\ref{eq:12}), for each of the three pairs of $U(1)$'s, and also a mixed equation (\ref{eq:111})  involving all three $U(1)$'s.
Each fermion may be charged under all three $U(1)$ groups, or under two of them, or even under a single  $U(1)$.

We will first show that all $U(1)_1\times U(1)_2\times U(1)_3$ theories with 7 or fewer fermions are vectorlike. 
Then we will give an explicit parametrization for some chiral solutions to the anomaly equations 
with 8 Weyl fermions charged under $U(1)_1\times U(1)_2\times U(1)_3$, \ie, $n_\chi=8$.
Finally, we will comment on theories with more $U(1)$ groups.

\subsection{$U(1)^3$ gauge theory with 7 Weyl fermions}

Let us search for $U(1)_1\times U(1)_2\times U(1)_3$ anomaly-free gauge theories with 7 Weyl fermions. 
The charges under $U(1)_j$ form the set $\{\vec{z}_j \}$, for $ j = 1,2,3$.
We first show that any chiral and anomaly-free $U(1)^m$ gauge theory with $m \ge 2$ must include for each $U(1)$ 
at least 4 fermions with nonzero charges under it. 
If there were fewer than four fermions, then the $U(1)$ and 
$[U(1)]^3$ anomalies would vanish only in the case of two charged fermions which form a vectorlike pair with respect to that  $U(1)$. In that case, however, the mixed anomalies force those two fermions to be vectorlike under the whole $U(1)^m$ group.

Next we will make a sequence of invertible transformations, as in Eq.~(\ref{eq:linear-comb-gen}), that will place two zero charges under each of the three $U(1)$'s for different fermions. This will allow us to show, using the anomaly equations, that the theory with 7 fermions is vectorlike. Since the linear transformation (\ref{eq:linear-comb-gen}) cannot transform chiral solutions into  vectorlike ones, this will prove that any chiral theory with $m =3$ requires more than 7 fermions.

First, use some nonzero elements of $\{\vec{z}_2\}$ and $\{\vec{z}_3\}$  to form a linear combination of charges 
that leads to two zeros in $\{\vec{z}_1\}$, and then sort the order of the fermions to get  $\{\vec{z}_1\} = \{ 0, 0 , z_{13}, z_{14}, z_{15}, z_{16}, z_{17} \}$.
Our earlier result that there are at least 4 fermions with nonzero charges under each $U(1)$ implies that the set  $\{\vec{z}_3\}$ has at least two nonzero charges which are not $z_{31}$ and $z_{32}$. Use one of these and one of the 
nonzero charges of $\{\vec{z}_1\}$ to put two zeros in $\{\vec{z}_2\}$. Again, sort the order of the fermions, but without changing the positions of the zeros in $\{\vec{z}_1\} $, to get $\{\vec{z}_2\} = \{ z_{21}, z_{22}, 0, 0, z_{25}, z_{26}, z_{27} \}$.  Finally, $\{\vec{z}_1\}$ and $\{\vec{z}_2\}$ have at least two elements different from zero besides $z_{21},z_{22},z_{13}$ and $z_{14}$. Therefore we can use a combination of these nonzero elements to put two zeros in $\{\vec{z}_3\}$  giving 
$\{\vec{z}_3\} = \{ z_{31}, z_{32}, z_{33}, z_{34}, 0, 0, z_{37} \}$.

Now, using the $U(1)_1 U(1)_2 U(1)_3$  anomaly equation, we find that at least one of $z_{17}$, $z_{27}$ and $z_{37}$ is zero. Without loss of generality, we take $z_{37}=0$, implying that $\{\vec{z}_3\}$ 
has exactly 4 nonzero entries. As a result, $\{\vec{z}_3\}$ includes two vectorlike pairs, so that there are only two possibilities
for the anomaly-free solutions with 7 fermions:   
\be   
\hspace{0.7cm}
 \!\!  \left[   \!\!
 \begin{array}{c|c|c}
 0 & \,  z_{21}  &  \  z_{31} \\
 0 & \, z_{22} &  \  - z_{31}  \\
 z_{13}  & \, 0  &  \  z_{33} \\
z_{14}   &  0 &  \    - z_{33}  \\
 z_{15}  &  z_{25}    &  \  0 \\
z_{16}    & z_{26}    &  \  0 \\
z_{17}    & z_{27}    &  \  0  \\
\end{array}    \! \right]     \hspace{1.1cm}   ,   \hspace{1.1cm}
 \!\!  \left[   \!\!
 \begin{array}{c|c|c}
 0 & \,  z_{21}  &  \  z_{31} \\
 0 & \, z_{22} &  \   z_{32}  \\
 z_{13}  & \, 0  &  \  - z_{31} \\
z_{14}   &  0 &  \    - z_{32}  \\
 z_{15}  &  z_{25}    &  \  0 \\
z_{16}    & z_{26}    &  \  0 \\
z_{17}    & z_{27}    &  \  0  \\
\end{array}    \! \right]       ~~.
\ee
The $U(1)_j [U(1)_3]^2$ and $[U(1)_j]^2 U(1)_3$ anomaly equations then imply $z_{14} = - z_{13}$ for $j = 1$,  and $z_{22} = - z_{21}$ for $j = 2$, as well as $z_{32} = - z_{31}$, so the  
 solution is vectorlike. We conclude that all $U(1)_1\times U(1)_2\times U(1)_3$ anomaly-free gauge theories with 7 or fewer Weyl fermions are vectorlike.

\bigskip

\subsection{Solutions for $U(1)^3$ anomaly equations with 8 fermions}

We now show that chiral solutions for the $U(1)_1\times U(1)_2 \times U(1)_3$ anomaly equations exist when there are 8 Weyl fermions.
We will not attempt to derive the general solution of this type.  Instead, we explicitly construct solutions which are vectorlike with respect to each $U(1)$ group taken separately, but are chiral with respect 
to the whole gauge group. Extending the terminology of Section \ref{sec:doublyvec}, these are triply-vectorlike solutions.

For the case of $U(1)_1\times U(1)_2$ we showed how to take any 6-fermion solution to the $U(1)$ anomaly equations and extend it to be a solution to the $U(1)_1\times U(1)_2$ anomaly equations.  Here, we will not start from the general 8-fermion solution for a single $U(1)$, generated by Eq.~(\ref{eq:even-merger}), but instead show how any $U(1)^3$ solution can be related to a combination of vectorlike solutions and the 6-fermion solution of Eq.~(\ref{eq:U(1)U(1)}).  
Starting from a solution to the $U(1)_1\times U(1)_2 \times U(1)_3$ anomaly equations we can always use linear combinations of charges, as in Eq.~(\ref{eq:linear-comb-gen}), to arrange for a pair of charges to be vectorlike across two of the $U(1)$ groups.  Without loss of generality we take this to be the last two charges and the first two groups.  That is, we combine $\{\vec{z}_3\}$ with $\{\vec{z}_1\}$ and $\{\vec{z}_3\}$ with $\{\vec{z}_2\}$ so that $z_{17}=-z_{18}$ and $z_{27}=-z_{28}$.  The first 6 charges by themselves must then solve all anomaly equations for the first two groups.  This means that any solution to the $U(1)_1\times U(1)_2 \times U(1)_3$ anomaly equations can be built by extending a solution to the $U(1)_1\times U(1)_2$ anomaly equations.  Furthermore, taking a linear combination of $\{\vec{z}_1\}$, $\{\vec{z}_2\}$, and $\{\vec{z}_3\}$ it is possible to make $\{\vec{z}_3\}$ purely vectorlike.  

The general solution for $U(1)_1\times U(1)_2 \times U(1)_3$ starts from a chiral 6-fermion solution to $U(1)_1\times U(1)_2$ and contains many possible orderings of vectorlike charges under $U(1)_3$.  
We leave the further investigation of the general solution for future work.  Here we focus on a less general case and construct a triply-vectorlike extension of the doubly-vectorlike solution for $U(1)_1\times U(1)_2$ presented in Section~\ref{sec:doublyvec}, selecting one of the many possible choices for the arrangements of vectorlike charges  $\{\vec{z}_3\}$.  We start from the first doubly-vectorlike solution given in Table~\ref{table:2}.  The solution has the structure,
\\
\be
\left[
\begin{array}{c|c|c}
j_1 &    j'_1       &   j''_1 \\
-j_1 &   j'_2     &   j''_2 \\
j_2 &    -j'_1     &   -j''_1 \\
-j_2 &     j'_3    &   -j''_2 \\
j_3 &      -j'_2   &  j''_3 \\
-j_3 &    -j'_3   &   j''_4 \\
j_4 &    j'_4   &   -j''_3 \\
-j_4 &   -j'_4   &   -j''_4\\
\end{array}
\right] ~,
\label{eq:triplevecansatz}
\ee
where $j'_4$ and $j''_1,\ldots,j''_4$ are integers constrained by the anomaly equations, while $j'_1, j'_2, j'_3$ are given by Eq.~(\ref{eq:quadratic-pol}).  The anomaly equations for $U(1)_1\times U(1)_2$ are automatically satisfied and there are multiple solutions to the five remaining mixed anomaly equations.  We have identified two simple chiral solutions of the $U(1)_1\times U(1)_2 \times U(1)_3$ anomaly equations with 8 fermions, each depending upon 3 integer parameters:
\be
\left[
\begin{array}{c|c|c}
j_1 & j'_1       &  j_3 \\
-j_1 & j'_2     & -j_3 \\
j_2 & -j'_1     & -j_3 \\
-j_2 &  j'_3    & j_3 \\
j_3 &   -j'_2   &  j_1 \\
-j_3 & -j'_3   & j_2 \\
-j_3 & -j'_1   & -j_1 \\
j_3 &   j'_1   & -j_2\\
\end{array}
\right] 
\quad\mathrm{and}\quad
\left[
\begin{array}{c|c|c}
j_1 & j'_1       &  -j_1+j_2+j_3 \\
-j_1 & j'_2     & j_3 \\
j_2  & -j'_1    & j_1-j_2-j_3 \\
-j_2 &  j'_3    & -j_3 \\
j_3  &    -j'_2   &  -j_1 \\
-j_3 & -j'_3   & -j_2 \\
-j_1+j_2+j_3 & -j'_1   & j_1 \\
j_1-j_2-j_3 &   j'_1   & j_2\\
\end{array}
\right] 
 ~~~,
\label{eq:tripleveccolutions}
\ee
where $j'_1, j'_2, j'_3$ are the quadratic polynomials (\ref{eq:quadratic-pol}) in the $j_1, j_2, j_3$ parameters.  This completes the proof that the smallest number of Weyl fermions charged under $U(1)^3$ that leads to chiral anomaly-free theories is $n_\chi=8$.  

Note that these solutions contain fermions which are vectorlike with respect to two of the $U(1)$'s but chiral with respect to all three.  These triply-vectorlike solutions can be made chiral under each $U(1)$ by taking linear combinations of charges, as in Eq.(\ref{eq:linear-comb-gen}).  As a numeric example consider the solution in Eq.~(\ref{eq:tripleveccolutions}) with $j_1=1, j_2=2, j_3=3$, put into canonical form,
\be
\left[
\begin{array}{c|c|c}
3 &   9     & 2\\
-3 &  -9    & 1 \\
3  &  -1    &  -1 \\
-3 &  7     &  -2 \\
2 &   -9     & 3 \\
-2 &  -7    & -3 \\
1 &    9     &  -3 \\
-1 &   1     & 3 \\
\end{array}
\right] 
\quad\quad\mathrm{and}\quad\quad
\left[
\begin{array}{c|c|c}
4 & 9   &1 \\
-4 &   -9   & 2\\
3  &    1   &  -1 \\
-3 & -7   & -2 \\
2  & 9    & -4 \\
-2 &  7    & -3 \\
1 & -9       &  4 \\
-1 & -1     & 3 \\
\end{array}
\right] ~.
\ee
After the linear combinations $\{\vec{z}_1\} \rightarrow \{\vec{z}_1\} +\{\vec{z}_2\}$, $\{\vec{z}_2\} \rightarrow \{\vec{z}_2\} +3\{\vec{z}_3\}$, $\{\vec{z}_3\} \rightarrow 2\{\vec{z}_3\} -\{\vec{z}_1\}+\{\vec{z}_2\}$ for the first solution, and $\{\vec{z}_1\} \rightarrow (\{\vec{z}_1\} -\{\vec{z}_2\}+2\{\vec{z}_3\})/3$, $\{\vec{z}_2\} \rightarrow \{\vec{z}_2\} +\{\vec{z}_3\}$, $\{\vec{z}_3\} \rightarrow \{\vec{z}_3\} +\{\vec{z}_1\}+\{\vec{z}_2\}$ for the second, these two triply-vectorlike solutions will transform to ones which are either composite or fully chiral under each $U(1)$,
\be
\left[
\begin{array}{c|c|c}
12 &   15   & 10\\
-12 & -6   & -4 \\
10 & 0       &  2 \\
-9 &  -16    & -11 \\
-7 & 0     & -5 \\
4 & 1   & 6 \\
2 &   -4   &  -6 \\
0 & 10     & 8 \\
\end{array}
\right] 
\quad\quad\mathrm{and}\quad\quad
\left[
\begin{array}{c|c|c}
6   &   5      &    4 \\
-5  &   -5    &   -7 \\
-5 &    -4    &   -2 \\
3  &    7   &     11\\
2  &   -2     &   -1 \\
-1 &   -10   &   -14 \\
0  &    9   &    12 \\
0  &    0   &    -3 \\
\end{array}
\right] ~,
\ee
where we have again reordered charges (and flipped the overall signs of some columns) 
to place them in canonical form.
In these examples, the individual $\{\vec{z}_i\}$ contain chiral sets with 5, 6, or 8 charges.

\subsection{More $U(1)$ gauge groups}

Consider the case of $U(1)^m$ gauge theories with $m \ge 4$.  As before, $n_\chi$ denotes the smallest number of fermions such that the theory is chiral and anomaly free, so the total number of charges is $n_\chi m$. Since the number of equations that these charges need to satisfy, ${\cal N}_{\rm eq}$ given in (\ref{eq:neq}),  grows 
as $m^3$, one could expect that $n_\chi$ will grow quadratically with $m$. It turns out that is not true: $n_\chi$ grows at most linearly with $m$.

To see that, notice that we can use the solutions we already have to construct composite solutions for larger $m$, placing the previous solutions in a 
block-diagonal way. For example, the $m=4$ case can be constructed with two $U(1)\times U(1)$ solutions with a total of 12 fermions, the first 6 fermions having nonzero charges under the first two $U(1)$'s, and the last 6 fermions having nonzero charges under the last two groups. 
Similarly, for $m=5$ we can use the $U(1)\times U(1)$ and $U(1)\times U(1)\times U(1)$ solutions, a total of $6+8=14$ Weyl fermions.  The zero entries in these block-diagonal solutions can be removed by taking linear combinations of charges, which changes the fermion charges but keeps the theory anomaly free.

For any $m$, using the solutions for three or two $U(1)$ groups
we can construct a solution with $8m/3$ fermions if $m = 0\!\mod 3$, or $8(m -1)/3 + 4$ fermions if $m = 1\!\mod 3$,
 or $8(m -2)/3 + 6$ fermions if $m = 2\!\mod 3$.  Thus, the number of Weyl fermions is linear in $m$ and not quadratic.
Chiral solutions which are not block diagonal may have even fewer fermions.

As an example, we construct a composite solution for $m=5$ starting from a 6-fermion chiral-vectorlike solution to $U(1)\times U(1)$ and an 8-fermion triply-vectorlike solution to $U(1)\times U(1)\times U(1)$.  In particular we use the chiral-vectorlike solution generated by Eqs.~(\ref{eq:6set}) and (\ref{eq:vset}) with $(k_1,\,k_2,\,\ell_1,\,\ell_2) = (2, 0, 1, -1)$ and the first triply-vectorlike solution of Eq.~(\ref{eq:tripleveccolutions}) with $(j_1,\,j_2,\,j_3) = (1,2,3)$.  By taking linear combinations of charges, the block diagonal structure is removed and we obtain the following set in canonical form:
\be
\left[
\begin{array}{c|c|c|c|c}
18 & 2 & 8 & 2 & 1 \\
-17 & -7 & -5 & -7 & 8 \\
-15 & 7 & -11 & 7 & -16 \\
12 & -8 & 10 & -8 & 17 \\
12 & 8 & 2 & 8 & -11 \\
-11 & -8 & -2 & 0 & -4 \\
10 & 13 & 4 & 6 & 2 \\
-10 & -2 & -4 & -2 & 1 \\
8 & -4 & -1 & -9 & 7 \\
6 & 3 & 12 & 10 & 14 \\
-6 & 6 & -3 & 5 & -11 \\
-5 & -8 & -14 & -16 & -12 \\
-4 & -16 & -7 & -15 & 1 \\
2 & 14 & 11 & 19 & 3 \\
\end{array}
\right] ~~.
\ee
Although no longer immediately apparent, this solution is still composite and consists of two subsets of six and eight fermions which are separately anomaly free.
\bigskip

\section{Conclusions} \setcounter{equation}{0}
\label{sec:conclusions}

The gauge charges of chiral fermions cannot be freely chosen.  The requirement of anomaly cancellation, dictated by a set of 
cubic Diophantine equations, places strong and complicated constraints on both the number and charges of the fermions that can be present in consistent theories.   
In this article we have analyzed the solutions to the anomaly equations for 
various Abelian gauge theories with the smallest number of Weyl 
fermions necessary to form anomaly-free chiral sets. 

For a single $U(1)$ gauge group the anomaly equations take the form ``the sum of the cubes equals the cube of the sum", and the most 
general solution was derived in \cite{Costa:2019zzy} for an arbitrary number of fermions. We have shown here that for the smallest chiral theory, which contains 5 Weyl fermions, the general solution has  different  properties than theories with more fermions: the charges are given by cubic polynomials in 3 integer parameters [see Eq.~(\ref{eq:5set})], the relative signs of the 5 charges are unique, and no two charges can be equal.  We also proved that there is a countably infinite number of such solutions.

For the $U(1) \times U(1)$ gauge theories, in addition to the anomalies for each $U(1)$, there are two mixed anomaly equations.
 We showed that at least 6 chiral fermions are necessary to satisfy all anomaly equations,  and we constructed the general $U(1) \times U(1)$ 
 solution for the 6-fermion set.  Remarkably, any anomaly-free chiral set of 6 charges under a single $U(1)$, for which the general solution is parametrized in Eq.~(\ref{eq:6set}), can be extended to be a solution with 6 chiral fermions charged under $U(1) \times U(1)$.  
 A set of fermions which is chiral under $U(1) \times U(1)$ may be doubly-chiral (\ie, chiral under each $U(1)$ group),
doubly-vectorlike  (\ie, vectorlike under each $U(1)$ group), or chiral-vectorlike.
We have found that the general doubly-chiral solution depends upon 6 integer parameters and a discrete choice from 6 arrangements (or branches) of the charges under the second $U(1)$, as shown in Eqs.~(\ref{eq:U(1)U(1)}) and (\ref{eq:U(1)U(1)uw}).  
All chiral-vectorlike solutions are particular cases of the general doubly-chiral solution, with the charges under the vectorlike $U(1)$ 
determined entirely by the charges under the chiral $U(1)$ [see Eqs.~(\ref{eq:v135}) and (\ref{eq:zeta0})].  The general doubly-vectorlike solutions depend upon 3 integer parameters and a discrete choice from 8 branches, as shown in Table \ref{table:2} and Eq.~(\ref{eq:quadratic-pol}). 
Finding the general solution for $U(1) \times  U(1)$ with $n\ge 7$ fermions is a challenging problem since the number of branches grows quickly with $n$, see Eq.~(\ref{eq:bp}).  

As the number $m$ of $U(1)$ gauge groups increases, the number of anomaly equations that must be satisfied grows as $m^3$.  
For the $U(1) \times  U(1) \times U(1)$ gauge theory, which has 13 anomaly equations, 
 we proved that the minimum number of chiral fermions is $n_\chi=8$,   
and we constructed a couple of 3-parameter chiral solutions, see Eq.~(\ref{eq:tripleveccolutions}).  For $U(1)^m$ with $m \ge 4$ we demonstrated how to build composite chiral solutions where the number of fermions grows linearly with $m$. Finding the general solution for $m\ge 3$ remains an unsolved problem.

The results and methods presented here have broad potential application, such as dark matter models, extensions of the Standard Model with $Z'$ bosons, or neutrino model building.  As an illustration, we presented a chiral dark matter model with $U(1) \times  U(1)$ gauge group and three dark fermions of hierarchical masses.

\bigskip\bigskip\bigskip

\noindent
{\bf  Acknowledgments:}  {  DC was supported by Funda\c{c}\~ao de Amparo \`a Pesquisa do Estado de S\~ao Paulo. 
BD and PF are supported by Fermi Research Alliance, LLC under Contract DE-AC02-07CH11359 with the U.S. Dept. of Energy.
}

\bigskip\bigskip


\vfil

\begin{thebibliography}{99}  
  
 
\bibitem{Bardeen:1969md} 
  S.~L.~Adler,
  ``Axial vector vertex in spinor electrodynamics,''
  Phys.\ Rev.\  {\bf 177}, 2426 (1969). \\
  W.~A.~Bardeen,
  ``Anomalous Ward identities in spinor field theories,''
  Phys.\ Rev.\  {\bf 184}, 1848 (1969). \\
  C.~Bouchiat, J.~Iliopoulos,  
  P.~Meyer,
  ``An anomaly-free  version of Weinberg's model,''
  Phys.\ Lett.\  B{\bf 38}, 519 (1972) \\
  D.~J.~Gross and R.~Jackiw,
  ``Effect of anomalies on quasi- renormalizable theories,''
  Phys.\ Rev.\ D{\bf 6}, 477 (1972).  \\
  H.~Georgi and S.~L.~Glashow,
  ``Gauge theories without anomalies,''
  Phys.\ Rev.\ D {\bf 6}, 429 (1972). 
  
\bibitem{Preskill:1990fr} 
  J.~Preskill,
  ``Gauge anomalies in an effective field theory,''
  Annals Phys.\  {\bf 210}, 323 (1991).
  
\bibitem{Banks:2010zn} 
  T.~Banks and N.~Seiberg,
  ``Symmetries and Strings in Field Theory and Gravity,''
  Phys.\ Rev.\ D {\bf 83}, 084019 (2011)
  [arXiv:1011.5120 [hep-th]].
  
\bibitem{Hilbert}
Y.~V.~Matijasevich,  ``Hilbert's Tenth Problem", Cambridge, MA: MIT Press, 1993.  \\
For a recent discussion, see  J.~Halverson, M.~Plesser, F.~Ruehle and J.~Tian,
  ``Kahler moduli stabilization and the propagation of decidability,''
  arXiv:1911.07835. 
  
\bibitem{Hardy}
G.~H.~Hardy and E.~M.~Wright,   ``An introduction to the theory of numbers", 
Oxford Univ. Press,  6th edition, 2008. 
 
\bibitem{Dickson}
L.~E.~Dickson, ``History of the Theory of Numbers", Vol. 2, Dover Publications, 2005.
  
\bibitem{Costa:2019zzy} 
  D.~B.~Costa, B.~A.~Dobrescu and P.~J.~Fox,
  ``General solution to the U(1) anomaly equations,''
  Phys.\ Rev.\ Lett.\  {\bf 123}, no. 15, 151601 (2019)
  [arXiv:1905.13729 [hep-th]].
   
\bibitem{Allanach:2018vjg} 
  B.~C.~Allanach, J.~Davighi, S.~Melville,
  ``An anomaly-free Atlas: charting the space of flavour-dependent gauged $U(1)$ extensions of the Standard Model,''
  JHEP {\bf 1902}, 082 (2019)
  [arXiv:1812.04602]. 
   
\bibitem{Batra:2005rh} 
  P.~Batra, B.~A.~Dobrescu and D.~Spivak,
  ``Anomaly-free sets of fermions,''
  J.\ Math.\ Phys.\  {\bf 47}, 082301 (2006)
  [hep-ph/0510181].
  
\bibitem{deGouvea:2015pea} 
  A.~de Gouvea and D.~Hernandez,
  ``New chiral fermions, a new gauge interaction, Dirac neutrinos, and dark matter,''
  JHEP {\bf 1510}, 046 (2015)
  [arXiv:1507.00916].   \\  
  J.~M.~Berryman, A.~de Gouvea, D.~Hernandez, 
   K.~J.~Kelly,
  ``Imperfect mirror copies of the standard model,''
  Phys.\ Rev.\ D {\bf 94}, no. 3, 035009 (2016)
  [arXiv:1605.03610]. 
 

\bibitem{Lu:2019rro} 
  Y.~Lu and J.~A.~Minahan,
  ``Notes on anomalies, elliptic curves and the BS-D conjecture,''
  arXiv:1908.04115 [hep-th].   \\
  J.~Rathsman and F.~Tellander,
  ``Anomaly-free model building with algebraic geometry,''
  Phys.\ Rev.\ D {\bf 100}, no. 5, 055032 (2019)
  [arXiv:1902.08529 [hep-ph]].
  
\bibitem{Babu:2003is} 
  K.~S.~Babu and G.~Seidl,
  ``Simple model for (3+2) neutrino oscillations,''
  Phys.\ Lett.\ B {\bf 591}, 127 (2004)
  [hep-ph/0312285].    
   
 \bibitem{Davoudiasl:2005ks} 
  H.~Davoudiasl, R.~Kitano, G.~D.~Kribs and H.~Murayama,
  ``Models of neutrino mass with a low cutoff scale,''
  Phys.\ Rev.\ D {\bf 71}, 113004 (2005)
  [hep-ph/0502176].
      
\bibitem{Appelquist:2002mw} 
  T.~Appelquist, B.~A.~Dobrescu and A.~R.~Hopper,
  ``Nonexotic Neutral Gauge Bosons,''
  Phys.\ Rev.\ D {\bf 68}, 035012 (2003)
  [hep-ph/0212073].
      
\bibitem{Carena:2004xs} 
  M.~Carena, A.~Daleo, B.~A.~Dobrescu and T.~M.~P.~Tait,
  ``$Z^\prime$ gauge bosons at the Tevatron,''
  Phys.\ Rev.\ D {\bf 70}, 093009 (2004)
  [hep-ph/0408098]. \\
  A.~Ordell, R.~Pasechnik, H.~Serôdio and F.~Teichmann,
  ``Classification of anomaly-free 2HDMs with a gauged U(1)' symmetry,''
  arXiv:1909.05548 [hep-ph].  \\
  B.~C.~Allanach, B.~Gripaios and J.~Tooby-Smith,
  ``Local anomalies in $Z^\prime$ models,''
  arXiv:1912.10022 [hep-th].
        
\bibitem{Bonnefoy:2019lsn}
For a few of the recent articles, see: \\ 
  Q.~Bonnefoy, E.~Dudas and S.~Pokorski,
  ``Chiral Froggatt-Nielsen models, gauge anomalies and flavourful axions,''
  arXiv:1909.05336 [hep-ph].  \\
  R.~Bause, M.~Golz, G.~Hiller and A.~Tayduganov,
  ``The new physics reach of null tests with $D \to \pi \ell \ell$ and $D_s \to K \ell \ell $ Decays,''
  arXiv:1909.11108 [hep-ph].  \\
  A.~Smolkovic, M.~Tammaro and J.~Zupan,
  ``Anomaly free Froggatt-Nielsen models of flavor,''
  JHEP {\bf 1910}, 188 (2019)
  [arXiv:1907.10063 [hep-ph]]. \\
  W.~Altmannshofer, J.~Davighi and M.~Nardecchia,
  ``Gauging the accidental symmetries of the Standard Model, and implications for the flavour anomalies,''
  arXiv:1909.02021 [hep-ph]. \\
  J.~Ellis, M.~Fairbairn and P.~Tunney,
  ``Anomaly-free models for flavour anomalies,''
  Eur.\ Phys.\ J.\ C {\bf 78}, no. 3, 238 (2018)
  [arXiv:1705.03447 [hep-ph]].

  \bibitem{Sayre:2005yh} 
  J.~Sayre, S.~Wiesenfeldt and S.~Willenbrock,
  ``Sterile neutrinos and global symmetries,''
  Phys.\ Rev.\ D {\bf 72}, 015001 (2005)
  [hep-ph/0504198].  
  
\bibitem{Babu:2004mj} 
  K.~S.~Babu and G.~Seidl,
  ``Chiral gauge models for light sterile neutrinos,''
  Phys.\ Rev.\ D {\bf 70}, 113014 (2004)
  [hep-ph/0405197]. \\
  M.~C.~Chen, A.~de Gouvea and B.~A.~Dobrescu,
  ``Gauge Trimming of Neutrino Masses,''
  Phys.\ Rev.\ D {\bf 75}, 055009 (2007)
  [hep-ph/0612017].
       
  
  
\bibitem{Heeck:2012bz} 
  J.~Heeck and H.~Zhang,
  ``Exotic charges, multicomponent dark matter and light sterile neutrinos,''
  JHEP {\bf 1305}, 164 (2013)
  [arXiv:1211.0538 [hep-ph]]. \\
  J.~M.~Berryman, A.~de Gouvea, K.~J.~Kelly and Y.~Zhang,
  ``Dark matter and neutrino mass from the smallest non-Abelian chiral dark sector,''
  Phys.\ Rev.\ D {\bf 96}, no. 7, 075010 (2017)
  [arXiv:1706.02722 [hep-ph]]. 
  
   
\bibitem{Nakayama:2011dj} 
K.~Nakayama, F.~Takahashi, 
T.~T.~Yanagida,
  ``Number-Theory Dark Matter,''
  Phys.\ Lett.\ B {\bf 699}, 360 (2011)
  [arXiv:1102.4688]. \\
  A.~Ismail, W.~Y.~Keung, K.~H.~Tsao and J.~Unwin,
 ``Axial vector $Z^\prime$ and anomaly cancellation,''
  Nucl.\ Phys.\ B {\bf 918}, 220 (2017)
  [arXiv:1609.02188].  \\
  B.~Batell,
  ``Dark Discrete Gauge Symmetries,''
  Phys.\ Rev.\ D {\bf 83}, 035006 (2011)
  [arXiv:1007.0045]. \\
  Y.~Cui and F.~D'Eramo,
  ``Surprises from complete vector portal theories: New insights into the dark sector and its interplay with Higgs physics,''
  Phys.\ Rev.\ D {\bf 96}, no. 9, 095006 (2017)
  [arXiv:1705.03897]. \\
L.~M.~Cebola, D.~Emmanuel-Costa, R.~Gonzalez Felipe and C.~Simoes,
  ``Minimal anomaly-free chiral fermion sets and gauge coupling unification,''
  Phys.\ Rev.\ D {\bf 90}, no. 12, 125037 (2014)
  [arXiv:1409.0805].  \\
  M.~C.~Chen, D.~R.~T.~Jones, A.~Rajaraman and H.~B.~Yu,
 ``Fermion mass hierarchy and proton stability from non-anomalous $U(1)_F$ in SUSY SU(5),''
  Phys.\ Rev.\ D {\bf 78}, 015019 (2008)
  [arXiv:0801.0248].  \\
  P.~Ko and T.~Nomura,
  ``Phenomenology of dark matter in chiral $U(1)_X$ dark sector,''
  Phys.\ Rev.\ D {\bf 94}, no. 11, 115015 (2016)
  [arXiv:1607.06218].\\
  J.~Ellis, M.~Fairbairn and P.~Tunney,
  ``Phenomenological Constraints on Anomaly-Free Dark Matter Models,''
  arXiv:1807.02503 [hep-ph].


\bibitem{Eguchi:1980jx} 
  For a review, see T.~Eguchi, P.~B.~Gilkey and A.~J.~Hanson,
  ``Gravitation, gauge theories and differential geometry,''
  Phys.\ Rept.\  {\bf 66}, 213 (1980).

\bibitem{Allanach:2019gwp} 
  B.~C.~Allanach, B.~Gripaios and J.~Tooby-Smith,
  ``Comment on `General solution to the $U(1)$ anomaly equations',''
  arXiv:1912.04804 [hep-th].


\bibitem{oeis}
See sequence A053871 in the ``Online Encyclopedia of Integer Sequences",\\ \url{https://oeis.org/A053871}  
  
\end{thebibliography}
\end{document}